\documentclass[12pt,preprint]{aastex631}

\newcommand{\bdv}[1]{\mbox{\boldmath$#1$}}

\def\au{{\rm au}} 
 
\def\kms{{\rm km}\,{\rm s}^{-1}}
\def\masyr{{\rm mas}\,{\rm yr}^{-1}}
\def\kpc{{\rm kpc}}
\def\mas{{\rm mas}}

\def\muas{\mu{\rm as}}

\def\pc{{\rm pc}}

\def\rel{{\rm rel}}

\def\eff{{\rm eff}}

\def\hel{{\rm hel}}

\def\e{{\rm E}}
\def\bpi{{\bdv\pi}}
\def\bmu{{\bdv\mu}}

\def\btheta{{\bdv\theta}}

\def\bv{{\bf v}}

\begin{document}

\title{OGLE-2016-BLG-1195 AO: Lens, Companion to Lens or Source, or None of the Above?}

\author{Andrew Gould}
\affiliation{Max-Planck-Institute for Astronomy, K\"{o}nigstuhl 17,
69117 Heidelberg, Germany}
\affiliation{Department of Astronomy, Ohio State University, 140 W.
18th Ave., Columbus, OH 43210, USA}

\author{Yossi Shvartzvald}
\affiliation{Department of Particle Physics and Astrophysics, 
Weizmann Institute of Science, Rehovot 76100, Israel}

\author{Jiyuan Zhang}
\affiliation{Department of Astronomy,
Tsinghua University, Beijing 100084, China}

\author{Jennifer C. Yee}
\affiliation{Center for Astrophysics $|$ Harvard \& Smithsonian, 60 Garden
St., Cambridge, MA 02138, USA}

\author{Sebastiano Calchi Novati}
\affiliation{IPAC, Mail Code 100-22, Caltech, 1200 E. California Blvd.,
Pasadena, CA 91125}

\author{Weicheng Zang}
\affiliation{Department of Astronomy,
Tsinghua University, Beijing 100084, China}
\affiliation{Center for Astrophysics $|$ Harvard \& Smithsonian, 60 Garden
St., Cambridge, MA 02138, USA}

\author{Eran O. Ofek}
\affiliation{Department of Particle Physics and Astrophysics, 
Weizmann Institute of Science, Rehovot 76100, Israel}

\begin{abstract}

We systematically investigate the claim by \citet{ob161195c} to have
detected the host star of the low mass-ratio ($q<10^{-4}$)
microlensing planet OGLE-2016-BLG-1195Lb, via Keck adaptive optics (AO)
measurements $\Delta t=4.12\,$yr after the peak of the event ($t_0$).
If correct, this measurement would contradict the microlens parallax
measurement derived from {\it Spitzer} observations 
in solar orbit
taken near $t_0$.  We show that this host identification would be in
$4\,\sigma$ conflict with the original ground-based lens-source
relative proper-motion measurements. 
By contrast, \citet{gould22} estimated a
probability $p=10\%$ that the ``other star'' resolved by single-epoch
late-time AO would be a companion to the host or the microlensed source,
which is much more probable than a 4$\,\sigma$ statistical fluctuation.
In addition,
independent of this proper-motion discrepancy, the kinematics of this
host-identification are substantially less probable than those of the
{\it Spitzer} solution.  Hence, this identification should not be
accepted, pending additional observations that would either confirm or
contradict it, which could be taken in 2023.
Motivated by this tension, we present two additional investigations.
We explore the
possibility that \citet{ob161195c}      identified the wrong ``star''
(or stellar asterism) 
on which to conduct   their analysis.
We find that 
astrometry of KMT and Keck images favors a star (or asterism)
lying about 175 mas northwest of 
the one that they chose.
We also present event parameters from a combined fit to all survey data,
which yields, in particular, 
a more precise mass ratio, $q=(4.6\pm
0.4)\times 10^{-5}$.
Finally, we discuss the broader implications of minimizing such false
positives
for the first measurement of the planet mass function, which will
become possible when AO on next-generation telescopes are applied to
microlensing planets.

\end{abstract}


\section{{Introduction}
\label{sec:intro}}

The most systematically applicable method of measuring host masses for
microlensing planets is late-time imaging of the system when the host
and source have separated sufficiently to resolve them.  The only fundamental
requirements of the method are that the host be luminous and that the
planet/host mass ratio $q$ and Einstein timescale $t_\e$ had been adequately
measured from the original event.  If the host is luminous, the
measurement yields the heliocentric lens-source relative proper motion
$\bmu_{\rel,\hel}$ and the lens flux (in, e.g., the $K$ band) $K_L$.  By adjusting
$\bmu_{\rel,\hel}$ to the geocentric value $\bmu_\rel$, and combining this
with $t_\e$, one obtains the angular Einstein radius $\theta_\e = \mu_\rel t_\e$.
Then,
\begin{equation}
  \theta_\e \equiv \sqrt{\kappa M \pi_\rel}, \qquad
  \kappa\equiv {4 G\over c^2\au}\simeq 8.14\,{\mas\over M_\odot}
  \label{eqn:thetaedef}
\end{equation}
and
\begin{equation}
  K_L = M_K(M_{\rm host}) + 5\log{D_L\over 10\,\pc} + A_K(D_L),
  \label{eqn:klrelation}
\end{equation}
providing two relations between the host mass $M_{\rm host}$ and distance $D_L$,
which can then be solved for both quantities.
Then, the mass of the planet $m_p = q M_{\rm host}$
can be found from the known value of $q$.  Here,
$\pi_\rel\equiv \au(D_L^{-1} - D_S^{-1})$ is the lens-source relative parallax,
$M_K(M_{\rm host})$ is the absolute magnitude of the lens, and $A_K(D_L)$
is its extinction.

\citet{gould22} examined a wide range of issues associated with this
method, including degeneracies in converting from $\bmu_{\rel,\hel}$ to
$\bmu_\rel$ and in combining Equations~(\ref{eqn:thetaedef}) and
(\ref{eqn:klrelation}), uncertainties in various parameters such as
the mass-luminosity relation $M_K(M_{\rm host})$, the source parallax
$\pi_S$, the extinction $A_K(D_L)$, and the light-curve measurement of $t_\e$.

The most vexing issue identified by \citet{gould22} was that
the ``other object''
(i.e., other than the source) might not be the lens.  Rather it could
be either a companion to the lens, a companion to the source, or in rare
cases, an ambient star.  \citet{gould22} showed that such misidentifications
do not pose an issue of principle: they can almost always be resolved
by additional late-time observations.  In all non-pathological cases, if
the object is not the lens, the relative proper-motion vector derived
from the two late-time observations will not
be consistent with uniform motion and lens-source coincidence
at the time of the event.  From the discrepancy, one can usually distinguish
among these cases, and in particular if the object is a companion to the
source, then it will hardly move between epochs, thereby permitting the lens
to subsequently appear.

The main issue is that, in general, one may not know when to take a second
late-time observation.  Adaptive optics (AO) observations on large,
or extremely large telescopes (ELTs) are expensive, while \citet{gould22}
estimated that of order 150 mass measurements could be made at AO first
light on ELTs from the 2016-2022 planet detections by the Korean Microlensing
Telescope Network (KMTNet, \citealt{kmtnet}) alone.  Hence, he considered
the issue of the false positive rate to be crucial.

Motivated by these considerations, \citet{gould22} distinguished between
two cases: those with and without measurements of $\mu_\rel$ from the event
itself.  Such measurements require that the normalized source radius
$\rho=\theta_*/\theta_\e$ be measured, which in turn requires that the
source pass over a caustic or close to a cusp (and that these are covered
by the observations).  In this case, it is possible to determine
$\theta_\e=\theta_*/\rho$ and so $\mu_\rel = \theta_\e/t_\e$ from the
angular source radius $\theta_*$, which can be determined by standard
techniques \citep{ob03262}.

\citet{gould22} estimated that for about 1/4 of microlensing planets,
$\mu_\rel$ cannot be measured from the light curve
(i.e., finite-source effects are not detected).  For these, he estimated
that the false-positive rate (mainly from companions to the lens), would
be about 10\%.  Nominally, the false positive rate would be exactly
the same for events with $\mu_\rel$ measurements.  However, in
most cases, one would receive a warning from the
fact that this light-curve based $\mu_\rel$ was inconsistent with the
one derived from the late-time imaging.  \citet{gould22} estimated that
if these cases were excluded (because they could be scheduled for
additional late-time observations) then the false-positive rate could
be reduced to about 3\%.  With this precaution, the overall false-positive rate
would be about $(1/4)\times 10\% + (3/4)\times 3\% \sim 5\%$.  If this is
considered to be an acceptable level, then only about $(3/4)\times 7\%\sim 5\%$
of events would have to be subjected to additional late-time observations.
Otherwise, a more aggressive approach would be needed.

Recently, \citet{ob161195c} have adopted the opposite approach.  They imaged
the planetary microlensing event OGLE-2016-BLG-1195 and reported
a strong, $3\,\sigma$, disagreement with the $\mu_\rel$ measurements from the
discovery papers \citep{ob161195a,ob161195b}. None of the six previous
late-time imaging efforts, 
OGLE-2005-BLG-071 \citep{ob05071c},
OGLE-2005-BLG-169 \citep{ob05169bat,ob05169ben},
MOA-2007-BLG-400 \citep{mb07400b},
MOA-2009-BLG-319 \citep{mb09319b},
OGLE-2012-BLG-0950 \citep{ob120950b}, and
MOA-2013-BLG-220 \citep{mb13220b},
had yielded such a strong disagreement.  Nevertheless, rather than regarding
this disagreement as a warning sign that this was one of the $\sim 10\%$
expected rate of false positives, they assumed that the ``other star'' was
indeed the lens, that the two original light-curve-based
$\mu_\rel$ measurements were each in error by $3\,\sigma$,
and that these were superseded by their own measurement.

Here, we investigate these issues further.  We note that the two original
$\mu_\rel$ measurements were based on completely independent data sets and
were consistent with each other at the $1\,\sigma$ level.  We therefore
combine these two measurements in two ways.
First, we use the standard method of combining
independent measurements, and, second, we combine the two data sets at
the light-curve level.  On this basis, we further refine the conflict
between the light-curve and late-time-imaging determinations.  We then
propose a test to determine whether the \citet{ob161195c} identification
is indeed a false positive.

\section{{$\mu_\rel$ Tension I:  Comparison to Two Independent Light-curve Analyses}
  \label{sec:tens1}}

In their Table~2,
\citet{ob161195c} reported the offset between the source and another star
 (which they identified as the lens) to
be $\Delta\theta = 54.49\pm 2.70\,\mas$ at $t_0+\Delta t$, where
$\Delta t = 4.12\,$yr and
$t_0$ is the peak of the event, when the lens and source were separated
by $\sim 15\,\muas$, i.e., far too small to be of interest here.  If the
``other star'' is indeed the lens, then (ignoring for the moment the
lens-source relative parallactic motion -- see next paragraph),
this would correspond to a
heliocentric relative proper motion,
\begin{equation}
\mu_{\rel,\hel} = {\Delta\theta\over\Delta t} = 13.22\pm 0.66\,\masyr.
  \label{eqn:helpm}
\end{equation}
Before continuing, we note that \citet{ob161195c} incorrectly
report an error of $\sigma=0.89\,\masyr$ in their Table~2, evidently
by adding in quadrature the errors of the two components of $\bmu_{\rel,\hel}$.
During the 4.12 (equally, 0.12) years of elapsed time, Earth moved
approximately East by about 0.7 AU, causing the lens to appear
to move west by $\delta\theta= 0.7\,\pi_\rel \sim 90\,\muas(\pi_\rel/130\,\muas)$,
where we have normalized to the \citet{ob161195b} value, which
\citet{ob161195c} argue, based on their measurement, is way too large.
Even if the \citet{ob161195b} value is correct, and taking account
of the direction of the offset, $-32^\circ$, north through east,
this would affect the proper motion determination by only about
$0.01\,\masyr$, substantially more than an order of magnitude below the
measurement errors.  Hence, this effect can safely be ignored.

We now argue that this measurement is in $4\,\sigma$ disagreement
with the $\mu_\rel$ measurements made by \citet{ob161195a} and
\citet{ob161195b}, via the relation
\begin{equation}
\mu_\rel = {\theta_*\over t_*} = {\theta_*\over \rho t_\e},
  \label{eqn:geopm}
\end{equation}
where $\theta_*$ is the angular radius of the source, $\rho=\theta_*/\theta_\e$,
and $t_*$ is the source self-crossing time.  Substantially different
methods are used to measure $\theta_*$ and $t_*$.  Hence, we treat them
separately.  We note that these two measurements are virtually uncorrelated.

\subsection{{$t_*$ Measurement}
\label{sec:tstar}}

\citet{ob161195a} and \citet{ob161195b} analyzed disjoint photometric
data sets, and each fit these to planetary models, which automatically
yielded estimates of $t_*$.  In principle, one might simply take these
two measurements and combine them in the standard way to obtain the
best overall estimate.  However, for several reasons, we adopt a more
comprehensive approach.

Our main concern is to check that the overall fits presented in these
two papers are consistent.  If they were not, it would be evidence that
one or the other of these two measurements were dominated by systematics,
in which case it would not be appropriate to combine them in the naive way.
Second, each group reported multiple models, i.e., two and eight models,
respectively.  Both groups reported two classes of models, which both
labeled ``wide'' and ``close'', but which are more accurately called
``inner'' and ``outer'' \citep{gaudi97,ob190960}.  That is, in the inner
model, the source passes inside the planetary caustic (over a ridge
between the planetary and central caustics), while in the outer model, it
passes outside the planetary wing of a central caustic.  In both cases,
these models (or groups of models) are indistinguishable at the $1\,\sigma$
level.  Therefore, we must determine whether these models make essentially
identical predictions for $t_*$.  Third, \citet{ob161195b} simultaneously
analyzed {\it Spitzer} data and so reported measurements of the
microlensing parallax vector $\bpi_\e$.  Such measurements are subject to
a well-known four-fold degeneracy \citep{refsdal66,gould94}, which is the
reason that \citet{ob161195b} reported 4 times more models.  Although
we are not directly concerned with the $\bpi_\e$ measurements in this
section, we must determine whether including $\bpi_\e$ in the fit
substantially impacts the parameters that we are interested in.

To conduct these investigations, we first reparameterize the fits reported
by the two papers in terms of ``invariants''.  That is, one usually
expresses the solutions to planetary microlensing events in terms of the
seven parameters, $(t_0,u_0,t_\e,\rho,q,\alpha,s)$, where $u_0$ is the
impact parameter (normalized to $\theta_\e$), $\alpha$ is the angle of
the source trajectory relative to the planet-host axis, and $s$ is the
planet-host separation (also normalized to $\theta_\e$).  However,
\citet{mb11293} showed that for planets detected at high magnification
the four quantities
$t_\eff \equiv u_0 t_\e$,
$t_* \equiv \rho t_\e$,
$t_q \equiv q t_\e$, and
$f_S t_\e$, are usually ``invariants'', i.e., the fractional errors in these
quantities are smaller than those naively inferred from the fractional
errors of the two factors because these are anti-correlated.  Here,
$f_S$ is the source flux, although we will not be making use of the $f_S t_\e$
invariant until Section~\ref{sec:thetastar}.

Thus, our 7 parameters are
$(t_0,t_\eff,t_\e,t_*,t_q,\alpha,s)$.  We infer best estimates of the
invariant parameters by taking the products of the two factors from the
published tables.  For the error estimates, we proceed as follows.
First, for the cases that asymmetric errors are reported
(all from \citealt{ob161195b}), we symmetrize them.  Second, for an invariant
parameter $\eta t_\e$ (i.e., the product of two parameters from the fit,
$\eta$ and $t_\e$), we adopt
$\sigma(\eta t_\e)=\eta t_\e\sqrt{[\sigma(\eta)/\eta]^2-[\sigma(t_\e)/t_\e]^2}$.
Note that \citet{ob161195a} already reported $t_*$, while \citet{ob161195b}
reported $\rho$ and $t_\e$ separately.

Our first step was to check whether the two models (or eight models)
were consistent among themselves, i.e., had essentially the same
values and errors (except for $s$).  Half of the eight
\citet{ob161195b} models had negative $t_\eff$ and $\alpha$, but we
just considered the absolute values of these quantities for this purpose.
We found that the error bars were all the same to within a few percent.
Furthermore, we found that the largest difference between models was
generally much smaller than the error bars.  For \citet{ob161195a}, the largest
difference was 30\% of the error bar (for $t_0$), while most were of
order 20\%.  For \citet{ob161195b}, the largest
difference was 45\% of the error bar (for $t_\e$ for one solution), while the
others were of order 10\% to 20\%.  Therefore, for each parameter,
we report, in Table~\ref{tab:combo}, the simple average of all solutions
(2 or 8) from each paper, except that we report the values of $s$ from
the two topologies separately.  In Column 6, we report the difference
between the two papers divided by the quadrature sum of their errors.
Under the assumption that the results reported from the two papers are
not dominated by systematics, we expect these differences to be
unit-variance Gaussian
distributed.  The first six parameters (i.e., excluding $s_{\rm inner}$ and
$s_{\rm outer}$) are essentially uncorrelated, implying that $\chi^2$ is just
the sum of the squares of the values in this column, i.e., $\chi^2=5.7$
for 6 degrees of freedom (dof).  The last two rows are essentially uncorrelated
with the others, but highly correlated with each other because
$\sqrt{s_{\rm inner}s_{\rm outer}}\simeq s^\dagger_+ \equiv
(\sqrt{4 + u_{\rm anom}^2} + u_{\rm anom})/2$, where
$u_{\rm anom} = t_\eff\csc\alpha/t_\e = 0.64\,{\rm day}/t_\e$.  If we add this
seventh dof and evaluate it at the mean of their absolute values, i.e., 1.35,
then $\chi^2=7.5$ for 7 dof.  In either case, this test constitutes strong
evidence against systematics in either analysis.

Hence, it is justified to combine them, which is done in Columns 7 and 8.
In particular, we find
\begin{equation}
t_* =  0.0324 \pm 0.0019\,{\rm day},
  \label{eqn:tstareval}
\end{equation}
i.e., a 5.9\% error.

\subsection{{$\theta_*$ Measurement}
  \label{sec:thetastar}}

As for the $t_*$ measurement, the two papers relied on completely
independent data sets for the measurement of $\theta_*$, which provides
a powerful consistency check.  On the other hand, both papers employed
the same overall method, which is subject to the same systematic errors.
Therefore, the two $\theta_*$ measurements cannot simply
be averaged together as was done for $t_*$, in Table~\ref{tab:combo}.

The basic method \citep{ob03262} is:
first, measure the offset of the source relative to the red clump,
$\Delta[(V-I),I] = [(V-I),I]_S - [(V-I),I]_{\rm cl}$, on a color-magnitude
diagram (CMD);
second, make use of the ``known'' dereddened position of the red clump,
$[(V-I),I]_{\rm cl,0}$ \citep{bensby13,nataf13},
to calculate the dereddened source values
$[(V-I),I]_{\rm s,0} = [(V-I),I]_{\rm cl,0} + \Delta[(V-I),I]$; and third,
use a color/surface-brightness relation to derive $\theta_*$ from
$[(V-I),I]_{\rm s,0}$.   While the first step is a straightforward measurement
with (usually) equally straightforward error estimation, the other two
steps are subject to systematic errors that are more difficult to
quantify.

\citet{ob161195a} and \citet{ob161195b} found
$\Delta[(V-I),I]_{\rm B+2017} = (-0.355\pm 0.021,3.369\pm 0.018)$ and
$\Delta[(V-I),I]_{\rm S+2017} = (-0.37\pm 0.03,3.40\pm 0.04)$, respectively.
Before continuing, we note that, of course, each group used its own
measurement of $I_S$ from its fit to the data.  Because $f_S t_\e$ is
an invariant, and because the two fits differed by
$t_{\e,\rm B+2017}/t_{\e,\rm S+2017}=1.0226$, these $I_S$ values would differ
(after calibrating to the same CMD) by 0.024 mag.  We put these
on the same system by adopting the combined $t_\e$ from Table~\ref{tab:combo},
which yields adjusted values
$\Delta[(V-I),I]_{\rm B+2017} = (-0.355\pm 0.021,3.349\pm 0.018)$ and
$\Delta[(V-I),I]_{\rm S+2017} = (-0.37\pm 0.03,3.404\pm 0.04)$.

In our view, the \citet{ob161195a} error bars are underestimated.
First, they report the error in $I_S$ itself as ``0.001'', although
their reported error in $t_\e$, combined with $f_S t_\e$ invariance,
implies that it is 0.027, which is substantially larger than their
reported total error.  In addition, based on our extensive experience,
including making all of the CMDs used for the \citet{bensby13} calibrations,
we do not believe that the clump can be centroided to the precision
given by \citet{ob161195a}.  We note that \citet{ob161195a} also report
a separate measurement based on OGLE-IV data, $\Delta(V-I)=-0.39\pm 0.03$.
We then adopt an average of the two values for $\Delta I$ and of the
three values for $\Delta(V-I)$, 
and we use the \citet{ob161195b} error bars, which basically reflect the
difficulty of centroiding the clump:
\begin{equation}
\Delta[(V-I),I]_{\rm adopted} = (-0.37\pm 0.03,3.38\pm 0.04).
  \label{eqn:deltavmii}
\end{equation}

As mentioned above, the next two steps require estimates of the
systematic errors.  For this line of sight, the ``known'' position of
the clump is $[(V-I),I]_{\rm cl,0}=(1.06,14.44)$.  Based on our experience
carrying out the \citet{bensby13} calibration, we estimate the clump
color error as $1.06\pm 0.03$.  The \citet{nataf13} clump magnitude measurement,
when combined with various stellar physics arguments, led to an estimate
for the Galactocentric distance of $R_0=8.1\,\kpc$, in remarkable agreement
with subsequent direct observations of SgrA*.  Therefore, we estimate the
systematic error in the clump magnitude as $14.44\pm 0.02$, and so find 
\begin{equation}
[(V-I),I]_{S,0} = (0.69\pm 0.04,17.82\pm 0.05)
  \label{eqn:smvii}
\end{equation}

As described above, the \citet{ob03262} method derives $\theta_*$ from
the dereddened color and magnitude using a color/surface brightness relation.
Usually, one employs such a relation that is calibrated
from angular diameter measurements.  However, as the source has
almost exactly the color of the Sun, we use its color and absolute magnitude,
$[(V-I),M_I]_\odot = (0.71,4.10)$, and its radius $R_\odot = 695,700\,$km,
as well as the differential color relation $d\ln\theta_*/d(V-I)= 0.966$
quoted by \citet{ob161195a} to obtain
$\theta_* = 0.822\pm 0.037\,\muas$.  Finally, we must account for the
fact that the source star has an unknown composition and so may have a
somewhat different surface brightness from the Sun at the same color.
To account for this, we add 2\% in quadrature to the error and finally
obtain,
\begin{equation}
\theta_* = 0.822\pm 0.041\,\muas. 
  \label{eqn:thetastareval}
\end{equation}
For comparison, \citet{ob161195a} derived $\theta_* = 0.856\pm 0.019\,\muas$,
while \citet{ob161195b} derived $\theta_* = 0.82\pm 0.07\,\muas$.

Combining Equations~(\ref{eqn:tstareval}) and (\ref{eqn:thetastareval})
yields
\begin{equation}
\mu_\rel = {\theta_*\over t_*} = 9.27\pm 0.72\,\masyr
  \label{eqn:mureleval}
\end{equation}
Ignoring for the moment the difference between heliocentric and geocentric
proper motions\footnote{Under the assumption that the AO measurement correctly
gives the host-source relative proper motion, which is appropriate for this
test, $\pi_\rel=(\mu_\rel t_\e)^2/\kappa M \rightarrow 27\,\muas$.  Then,
according to Equation~(\ref{eqn:convert}), below, the 
correction to the vector proper motion is
$\Delta\bmu_\rel = 0.17\,\masyr\bv_{\oplus,\perp}/v_{\oplus,\perp}$, so the
correction to the scalar proper motion is
$\Delta\bmu_\rel\cdot \bmu_{\rel,\hel}/\mu_{\rel,\hel} = -0.095\,\masyr$.
Hence, it is justified to ignore this effect, but if we included it,
it would increase the tension by about $0.1\,\sigma$.
}, Equations~(\ref{eqn:helpm}) and (\ref{eqn:mureleval}) differ
by $3.95\pm 0.98\,\masyr$, i.e., a $4.0\,\sigma$ discrepancy.

\section{{{\it Spitzer} $\bpi_\e$}
\label{sec:spitzerpie}}

\citet{gould22} also mentions contradictions between light-curve-based
measurements of the microlens parallax $\bpi_\e$ and the lens parameters
derived from the AO imaging.  The microlens parallax is defined by
\begin{equation}
\bpi_\e \equiv {\pi_\rel\over\theta_\e}\,{\bmu_\rel\over\mu_\rel}.
  \label{eqn:piedef}
\end{equation}
In principle, it can be measured from light-curve distortions generated
by Earth's annual motion \citep{gould92}, but because for most
events (and OGLE-2016-BLG-1195, in particular), $t_\e\ll\,$yr, this is
often impossible, and it is difficult in most other cases.

Nevertheless,
OGLE-2016-BLG-1195 was observed by {\it Spitzer} from solar orbit \citep{yee15},
and such observations can in principle measure $\bpi_\e$ even for very
short events \citep{refsdal66,gould94}.  However, such satellite-parallax
measurements are only straightforward if the satellite observations cover
both the rising and falling sides of the light curve.  Because of constraints
on {\it Spitzer} operations, observations could not be immediately triggered.
In particular, observations of OGLE-2016-BLG-1195 did not begin until 1.6
days after the ground-based peak.  Moreover, the total flux variation
was only about 2.5 flux units, whereas in several other cases, it has
been shown that {\it Spitzer} light curves show systematics at the level
of 0.5--1 flux unit.  Hence, in our view, results derived from cases with
few-flux-unit variations must be treated cautiously, but can still provide
valuable information.  For example, {\it Spitzer} observations of Kojima-1
\citep{kojima1c}
covered only the extreme falling wing of the light curve, with a flux
variation of only 5 units, yet it delivered precise parallax information
that has been independently confirmed by other techniques
\citep{kojima1a,kojima1b}.

However, rather than seeing the contradiction with previous $\bpi_\e$
measurements as a reason for caution and deeper investigation of their
own results, \citet{ob161195c} took this contradiction as ``proof'' that
{\it Spitzer} $\bpi_\e$ measurements derived from low flux variations are
unreliable.  By contrast, in our view, such a conclusion would only
be appropriate if the AO results were independently confirmed in some other
aspect, such as agreeing with the very precise, and multiply-confirmed
measurement of $\mu_\rel$.  In fact, the radical disagreement of the AO-based
and light-curve-based proper-motion measurements implies that the AO
observations cannot be used to cast doubt on the {\it Spitzer} $\bpi_\e$.
Thus, in the following section, we will retain an open mind regarding the
{\it Spitzer} $\bpi_\e$ measurement.

\section{{Other Scenarios}
\label{sec:other}}

In general, when a planet/host mass measurement is made based on a single
late-time AO observation, one should always consider the possibility that
the ``other star'' is not the lens, but rather is a companion to either
the lens or source or is an ambient star.  Moreover, as pointed out by
\citet{gould22}, these possibilities must be taken even more seriously
when the AO measurement appears to contradict previously known facts about
the event.  The contradiction that is expected to be most frequent is between
the AO-based $\bmu_{\rel,\hel}$ and the light-curve-based $\mu_\rel$.  In the
present case, this contradiction is quite severe.  Moreover, as just
discussed in Section~\ref{sec:spitzerpie}, it is augmented
by a conflict between the AO measurement and the {\it Spitzer} $\bpi_\e$.
See Figure~3 of \citet{ob161195c}.

\subsection{{Companion to the Lens (Host)}
\label{sec:lens_comp}}

The ``other star'' could be a companion to the lens (host).  As we will soon
show, such a companion must be separated from the lens by at least
tens of $\theta_\e$ (so tens of astronomical units), and therefore
it will be moving with nearly identical proper motion to the lens.
Thus, it can be robustly predicted that a second epoch, which could be taken
``now'' (in 2023) will show a vector displacement from its 2020 position
of about $(2023-2020)\mu_\rel\simeq 28\,\mas$.  Such an observation would
rule out a source companion (which would hardly move) and would render the
explanation of an ambient star extremely unlikely.  Hence, it would confirm
the ``lens-companion'' hypothesis.  Furthermore, by extrapolating the
companion proper motion back to $t_0$ (in 2016) one would find the
separation between the ``lens'' (planet host) and its companion in arcsec.

However, such a measurement would not, in itself, tell us the mass or
distance of the host or the planet.  That is, the lens could be anywhere
along the line of sight, and at each possible distance, the lens would
have a mass that is consistent with the measured
$\theta_\e=\sqrt{\kappa M \pi_\rel}$, and the companion would have a mass
consistent with its measured flux.  Nevertheless, such a measurement
would tell us the future positions as functions of time
of the lens relative to both the source and the companion, which would permit
an informed decision on when the lens could be imaged (possibly with more
advanced instruments).  Even if the lens were dark, one could still determine
the distance of the companion (and so the lens system) by multi-color, or
possibly spectral observations of the companion.  Combining this distance
with the $\theta_\e$ measurement would then yield the host (and planet)
mass.  Thus, the first step is simply to obtain
another AO epoch, which could be done immediately.

Next, is the lens-companion hypothesis consistent with the {\it Spitzer}
$\bpi_\e$ measurement being correct?  Recall first that there were 8 such
$\bpi_\e$ solutions.  However, these come in 4 (inner/outer) pairs, whose
$\bpi_\e$ are nearly identical.  Moreover, the scalar amplitudes are very
similar among these four solutions: $\pi_\e=(0.437,0.473,0.482,0.430)$,
leading to very similar $\pi_\rel = (0.111,0.120,0.122,0.109)\,\mas$.
Hence, all solutions are at similar distances $D_L=4.2\,\kpc$, and they
also have similar lens masses $M_{\rm host}\sim 0.07\,M_\odot$.
At this distance, the measured flux of the companion $K=19.96$, together
with extinction $A_K=0.24$ \citep{gonzalez12} and the mass-luminosity
relation of \citet{benedict16}, yields a companion mass $M_{\rm comp}=0.4\,M_\odot$,
hence, a mass ratio\footnote{We use an upper case ``$Q$'' to distinguish
this putative stellar companion from the planet, whose mass ratio is
designated ``$q$''.} $Q = M_{\rm comp}/M_{\rm host} \simeq 6$.

In principle, such a massive companion could be ruled out because it
might predict a light-curve distortion near the peak of this relatively
high-magnification event that is not seen.  To determine whether this is
the case, we predict the position of the lens relative to the source,
using the measured $\mu_\rel$ and the four values of $\bpi_\rel$, to obtain
the corresponding $\bmu_\rel = \mu_\rel\bpi_\e/\pi_\e$, and finally convert
to heliocentric
\begin{equation}
\bmu_{\rel,\hel} = \bmu_\rel + {\pi_\rel\over\au}\bv_{\oplus,\perp},
  \label{eqn:convert}
\end{equation}
where $\bv_{\oplus,\perp}(N,E) = (-0.76,+28.94)\,\kms$, is the velocity of
Earth, projected on the sky at $t_0$.

\begin{figure}
\plotone{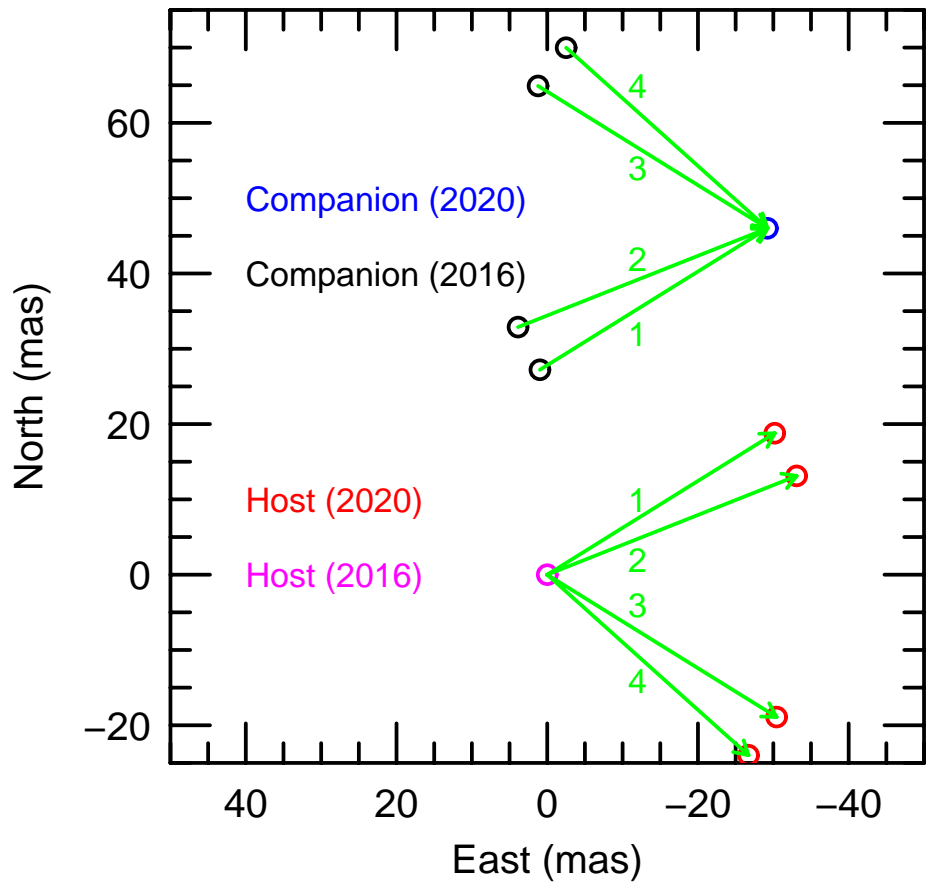}
\caption{Illustration of 4 models (corresponding to 4 degenerate solutions
for the microlens parallax $\bpi_\e$), in which the ``other star'' is
assumed to be a companion to the host.  The green arrows show the
vector motion over 4.12 years relative to the source (at the origin)
for the companion (upper part of diagram) and host (lower part of
diagram) for the four solutions, which are labeled sequentially by
decreasing $\pi_{\e,N}$.  In all cases, the host is superposed with the source
in 2016, while the companion is at the location measured by \citet{ob161195c}
in 2020.
}
\label{fig:motion}
\end{figure}

These models are illustrated in Figure~\ref{fig:motion}.  The green
arrows represent the four heliocentric proper motions, propagated over
the 4.12 years between $t_0$ and the Keck observations.  The blue point
is the measured position of the ``other star'' in 2020 relative to the
source, which is at the origin.  In these models the ``other star'' is
assumed to be a companion to the host.  The black points then represent
the companion position relative to the source (and so the host) at $t_0$.
The red points represent the position of the host relative to the source
in 2020.

Note that for two of the solutions, the companion was separated from the
source at $t_0$ by 65 or 70 mas, so if there had been an AO observation
at that time, such models could have been confirmed or ruled out.
However, to the best of our knowledge, no such observations were taken.
There were AO observations taken in 2018 by \citet{ob161195c}, when
the companion would have been separated from the source by about 57 or 60 mas
for these two models.  These observations were of lower quality than the 2020
observations, so it is not clear whether they could have detected the
companion at the position predicted by these two models (i.e., models
3 and 4 in Figure~\ref{fig:motion}).  In any case, there are no
constraints on models 1 and 2.

Note that models (1,2,3,4) predict separations between the lens companion
and the source of about $(56,60,76,78)\,\mas$ in mid 2023.  Hence, all would be
detectable under good conditions, while models 3 and 4 (the two solutions
that could not have been probed by the 2018 observations) would be detectable
even under moderately good conditions.

For all four solutions, the companion is separated from the lens
by at least $\Delta\theta_{\rm comp}>27\,\mas$, which corresponds to
$s_{\rm comp} = \Delta\theta_{\rm comp}/\theta_\e>106$.  This would induce
a \citet{cr1,cr2} caustic of radius $w=2Q/s_{\rm comp}^2\sim 10^{-3}$,
which is about 50 times smaller than the closest passage of the source, i.e.,
$u_0\sim 0.05$.  Hence, the companion would not have induced any
noticeable effect on the light curve near peak.

\subsection{{Companion to the Source}
  \label{sec:source_comp}}

The ``other star'' could be a companion to the source.  If so, at
$D_S\sim 8\,\kpc$ and $A_K=0.24$, it would have
$M_K\simeq 5.2$ and so $M= 0.6\,M_\odot$,
with projected separation $a_{\perp,S}\simeq\,435\,\au$.  According to
Figure~7 from \citet{dm91}, the corresponding period,
$\log (P/\rm day)\sim 6.9$, is within the broad peak of the distribution
for companions to solar-mass stars.  Similarly, companions of mass ratio
$\sim 0.6$ are also relatively common according to their Table~7.  Thus,
there is no reason to discount this possibility.  AO observations in 2023
could confirm this hypothesis, provided that they were taken under
similarly good conditions as the 2020 observation because the ``other star''
would remain at $\Delta\theta=54\,\mas$.  Note that two of the four
lens-companion models (under the assumption that the {\it Spitzer} $\bpi_\e$
measurement is correct) make similar predictions for $\Delta\theta$
(see Section~\ref{sec:lens_comp}), but these have very different position
angles (see Figure~\ref{fig:motion}).  Note also that even if conditions
are less than ideal, these observations would easily detect the ``other star''
at $\Delta\theta=92\,\mas$ under the \citet{ob161195c} hypothesis that
the ``other star'' is the host.  Note finally that if the ``other star'' is a
companion to the source, it would lie well outside the Einstein ring and
so would not be magnified at all during the event.

\subsection{{Ambient Star}
  \label{sec:ambient}}

The only other logical possibility (apart from host, companion to the host, and
companion to the source) is that the ``other star'' is an ambient star
that is unrelated to the event.  In this case, the star would have
$I_0\sim 21.2$.  The surface density of stars within a magnitude of this
value toward Baade's Window is about 1000/arcmin$^2$ \citep{holtzman98},
while the surface density toward OGLE-2016-BLG-1195 is twice that of
Baade's Window \citep{nataf13}.  Therefore, the expected number of
such stars within 54 mas is $p=5\times 10^{-3}$.  While small, it is still
much larger than the Gaussian probability that the light-curve based
proper-motion measurement is in error by $4\,\sigma$ ($p< 10^{-4}$).
Hence, it must be considered.

An additional AO epoch, 
of the ``other star'' could confirm the ambient-star hypothesis provided
that the resulting vector-proper-motion measurement meets the following
conditions: (1) it conflicts with the vector proper motion derived by
\citet{ob161195c} under the assumption that this star is the host; (2) 
it also conflicts with the scalar proper motion derived from the light curve
(which should apply to either the host or a companion to the host); and (3)
it is inconsistent with zero (as would be expected for a companion
to the source).

.\section{{Two Other Issues}
\label{sec:issues}}

There are two additional issues that impact the plausibility of
the identification of the ``other star'' as the lens.  While neither appears
to be as severe as the $4\sigma$ discrepancy in the proper motion measurement,
both do need to be considered.

\subsection{{Galactic Kinematics}
\label{sec:kinematics}}

One reason for concern about the original \citet{ob161195b} {\it Spitzer}
parallax measurement is that it appeared to imply that the kinematics
of the lens are strongly at variance with what is expected for stars at
$\sim 4\,\kpc$ within the Galactic disk.
That is, using our refined measurements
of $\mu_\rel$ and $t_\e$ (which are only slightly different from theirs)
and their measurements of $\bpi_\e$, we find for the least retrograde
solution (``solution 1'' in Figure~\ref{fig:motion}), that
$\bmu_{\rel,\hel}(N,E) = (+4.56,-7.33)\,\masyr$, which corresponds to
$\bmu_{\rel,\hel}(l,b) = (+0.25,+8.63)\,\masyr$.

For comparison, if the source had the mean motion of bulge stars as measured
by {\it Gaia} \citep{gaia16,gaia18}
toward this direction, $\bmu_{S,\hel}(l,b) = (-5.74,-0.13)\,\masyr$,
this would imply lens motion, $\bmu_{L,\hel} = (-5.49,+8.50)\,\masyr$.
By contrast, taking account of the motion of the Sun relative to the LSR,
and adopting an asymmetric drift at 4.2 kpc of $-25\,\kms$, the mean
motion of a disk lens at this distance would be
$\bmu_{\hel,\rm meanDisk}(l,b)=(-1.86,-0.35)\,\masyr$.
Of course, neither the source
nor the lens can be expected to be moving exactly at the mean motion of their
respective populations, but the difference,
$\Delta \bmu_{L,\hel}(l,b)=\bmu_{L,\hel}-\bmu_{\hel,\rm meanDisk}(l,b)=(-3.63,+8.85)\,
\masyr$,
i.e., a total of
$|\Delta \bmu_{L,\hel}|= 9.6\,\masyr$, would require peculiar motions of one
or both of these stars that are relatively rare.  That is, the dispersions
of bulge sources are only $(3.0,2.7)\,\masyr$ in the $l$ and $b$ directions,
while the dispersions of the disk lens at this distance are about
$(3.1,2.1)\,\masyr$.  Hence, the offset comes to
$(1.0,2.6)\,\sigma$ in the two directions and thus a 
probability $p=2\%$.  The implausibility of this
scenario was undoubtedly one of the motivations that led \citet{ob161195c}
to begin observing OGLE-2016-BLG-1195 in 2018, when the lens and source
were expected to be separated by only $\sim 18\,\mas$, meaning that
it was impossible to resolve them using Keck.

However, the identification of the ``other star'' as the lens does not resolve
this motivating issue.  Repeating the same steps as above, but assuming an
asymmetric drift of $-40\,\kms$ (at $D_L\sim 7\,\kpc$), we obtain
$\bmu_{\rel,\hel}(N,E) = (+11.16,-7.09)\,\masyr$, 
$\bmu_{\rel,\hel}(l,b) = (+6.16,+11.70)\,\masyr$,
$\bmu_{L,\hel} = (+0.42,+11.57)\,\masyr$,
$\bmu_{\hel,\rm meanDisk}(l,b)=(-1.59,-0.21)\,\masyr$,
$\Delta \bmu_{L,\hel}(l,b)=(+2.01,+11.78)\,\masyr$, and 
$|\Delta \bmu_{L,\hel}|= 11.95\,\masyr$.  At this distance, the disk
proper-motion dispersions
are similar, so that the kinematics of this
``solution'' are intrinsically less probable, $p=0.24\%$, than those of
``bad {\it Spitzer} measurement'' that it was intended to solve.

\subsection{{Apparent Source Star is Too Bright}
  \label{sec:bright}}

Another issue is that flux from the apparent source star.  That is, after
\citet{ob161195c} subtracted the flux of the ``other star'',
the remaining flux that they attribute to the source star
($K=16.98\pm 0.05$), is substantially higher than one would predict based on
models of the microlensing event.  Our best estimate of the intrinsic
source color and magnitude (Equation~(\ref{eqn:smvii})) combined with
the color-color relations of \citet{bb88}, would imply
$K_{S,0}= 17.02\pm 0.08$.  Adopting $A_K= 0.24$ \citep{gonzalez12}, this
implies $K_S= 17.26\pm 0.08$.  Hence, the apparent source is
$\Delta K = 0.28\pm 0.10$ mag brighter than expected.

\citet{ob161195c} do not comment on this discrepancy, but
there are a number of possible explanations for it,
some inconsequential but others that would substantially impact the
interpretation of the lens system.

In the absence of any other information, by far the most likely explanation
would be that the excess light is due to the lens.  Indeed, this was the
tentative conclusion of the Keck team when they first detected the excess light
during the analysis of their 2018 observations using the NIRC2 camera on Keck
(J.-P.\ Beaulieu, private communication 2019).  These observations, made when
the lens-source separation was only $\sim 20\,\mas$ could not possibly have
resolved the lens from the source, regardless of its brightness.  Hence,
the main value of these observations would be to detect excess light, or at
the least, serve as a first AO epoch that could help clarify later AO
observations.  Nevertheless, the Keck team ultimately adopted the cautious
approach of waiting for confirmation by subsequent AO observations.

In fact, it is exactly their 2020 Keck OSIRIS observations that rule out
the lens interpretation of this excess light: if the source/lens flux ratio
really were 3:1, and if they were separated
by $\sim 38\,\mas$ (as predicted by the microlensing model,
Equation~(\ref{eqn:mureleval})),
then the lens would have been resolved by exactly the same technique as was used
by \citet{ob161195c} to detect the much fainter object at $54\,\mas$.
For example, \citet{ob120950b} securely resolved the lens of OGLE-2012-BLG-0950
at $\Delta\theta\sim 34\,\mas$, with a flux ration 1.46:1.  See their Figure~3.

Two other possibilities are that this excess light is due to a companion
to the lens or to the source.  Broadly, these scenarios are similar
to those discussed in Sections~\ref{sec:lens_comp} and \ref{sec:source_comp},
so we do not discuss them in detail here.  The main difficulty is that,
if one takes the $4\,\sigma$ proper-motion discrepancy at face value,
then both this excess light and the ``other star'' reported by
\citet{ob161195c} are due to companions, without the lens yet being detected.
Nevertheless, this is certainly possible in principle.

Another possibility is that ``$3\,\sigma$ errors happen''.  That is,
there is in reality no excess light: the problem is an incorrect measurement
of the star's flux or a misestimate of the source color and magnitude from
the analysis of the microlensing event.  However, this $3\,\sigma$ error
is occurring on the back of another, independent, $4\,\sigma$ error, which
begins to strain credulity.

Finally, it is also possible that \citet{ob161195c} have identified the
wrong star as the (previously) microlensed source, and the actual
source is the neighbor that lies $\sim 175\,\mas$ to the northwest, at a
position angle, $\sim -31^\circ$.  This would be a very unsatisfying ``solution''
because it creates as many problems as it solves, but it does require
investigation.

Under this hypothesis, the ``other star'' would almost certainly be a
companion to its brighter neighbor at 54 mas, just as was considered  in
Section~\ref{sec:source_comp}.   Then the microlensed source would be
one component of an asterism that is several tenths brighter than the
star identified by \citet{ob161195c}, i.e., with excess light almost
equal to the microlensed source (instead of a 1:3 ratio).  Thus, this
scenario trades the problem of explaining a moderate amount of excess light
for a larger amount of excess light.  Again, in the context of this
hypothesis, the excess light could be the lens or a companion to the
lens or source.  The one difference from the original scenario is that
the lens explanation cannot (yet) be ruled out because no one has
applied the techniques of \citet{ob120950b} and \citet{ob161195c} to this
star.

Have \citet{ob161195c} misidentified the (formerly) microlensed source?
We cannot say with certainty.  They do not discuss how they made their
identification, except to note that they had already done so based on
their 2018 NIRC2 observations.  And they certainly did so using image-level
microlensing data that are not available to us.  However, we have carried
out our own determination using KMT image-level data, which we report
in Appendix~\ref{sec:sourcepos}.
In brief, we find that in 2018 the ``northwest star'' (lying about 175 mas from
the ``southeast star'' identified by \citealt{ob161195c})
lies roughly $30\pm 15\,\mas$ west of the microlensed source
(as determined in 2016), while in 2020, it lies roughly $45\pm 15\,\mas$ west
of the microlensed source. According to this analysis, the ``northwest star''
is a substantially better candidate for the source position.
Nevertheless, in the absence of an account of the
\citet{ob161195c} determination, we do not regard the matter as resolved.

\section{{$\mu_\rel$ Tension II:  Reanalysis of the Original Light-curve}
  \label{sec:tens2}}

As discussed in Section~\ref{sec:tens1}, the first response to an
apparent conflict between the light-curve based measurement of $\mu_\rel$
and the AO-based measurement of $\bmu_{\rel,\hel}$ should be a review of
the published literature to determine how secure this conflict actually
is.  We carried such a review and found the conflict to be at the
$4.0\,\sigma$ level.  The strength of this conflict provided the context
for investigating other possible explanations for the detection of the
``other star'' in Sections~\ref{sec:other} and \ref{sec:issues}.

However, another possibility is that the original light-curve analysis
was incorrect and that a corrected value of $\mu_\rel$ might be more
consistent with the AO-based $\bmu_{\rel,\hel}$.  In Section~\ref{sec:tens1},
we argued that the original analyses were likely to be robust because
there had been two such analyses that used independent data sets and
independent codes, and we measured the difference between these as
having $\chi^2/{\rm dof}=7.5/7$.

Nevertheless, there are several reasons for pursuing this course.
The most important is that
the analysis of even seemingly ``simple'' planetary events like
OGLE-2016-BLG-1195, has progressed substantially over the intervening 7
years and continues to do so.  In particular, new degeneracies have
been discovered and new methods for finding degenerate solutions are
being developed.  As we will review below, one of these degeneracies
can, in principle,
lead to dramatic changes in $t_*$ and therefore in $\mu_\rel=\theta_*/t_*$.
Second, while seemingly unlikely, it remains possible that mistakes
were made in the original analyses.  Third, the disagreement between
the two measurements of $t_*$ was at the $1.2\,\sigma$ level, with
the weighted average strongly dominated by the slower-$\mu_\rel$ result
of \citet{ob161195a}.  Hence, if there were an error in that analysis,
this could, by itself, significantly reduce the tension.  Finally,
there have recently been important improvements to the KMT data reductions,
and these could in principle also change the result.

Moreover, \citet{mb13220b} charted a path of such light-curve reanalyses
when they found that their AO observations conflicted with the
upper-limit on lens light derived from the original light-curve analysis
by \citet{mb13220} of MOA-2013-BLG-220\footnote{In their
initial arXiv preprint, their revised
fit showed a reduction of the Einstein timescale by $\Delta\ln t_\e = -0.147$,
and corresponding increase in source flux by $\Delta\ln f_S = +0.144$.
This would have substantially increased the tension by leaving even less of
the baseline flux available for lens light.  However, in the final
published version of the paper, they
corrected this fit and essentially reproduced the \citet{mb13220} fit
parameters.  They attributed the modest remaining discrepancy to
the effects of faint, unmodeled background stars.  In any case,
this appears to be the first effort to systematically track down
such discrepancies by checking the original light-curve models.
}.

While we mainly report in this section on the results using re-reduced KMT
data, we note that we first checked on the result of combining all the
light curves (i.e., OGLE, MOA, and KMT) from the published literature
\citep{ob161195a,ob161195b}.  We found that the best-fit values of all 8
parameters listed Table~\ref{tab:combo} were well within
the $1\,\sigma$ range shown in the final two columns of that Table.
By far, the largest ``discrepancy'' was for $t_0$, which differed by
  $\sim 0.5\,\sigma$.
In addition, we found that while the central value of $t_\eff$ was almost
identical to that of Table~\ref{tab:combo}, its error bar was only half as
big.

As stated above, to conduct our reanalysis, we first re-reduced the
KMT data using an improved pipeline (Yang et al., in prep)
that is now routinely applied to essentially all new published KMT events.
Inspection of the images showed that the KMTC01 data are affected by
a ``spike'' from a neighboring bright star.  Hence, we eliminated these
from the analysis.  Because KMTC does not cover the anomaly and because
(by favorable chance) the event lies in two other high-cadence fields
(BLG41 and BLG42),
the elimination of these data is not expected to have a major effect.

The results of our analysis are given in Table~\ref{tab:wz}, where we show
both the standard parameters and the three invariant parameters
$(t_\eff,t_*,t_q)$ in order to enable comparison with Table~\ref{tab:combo}.

\begin{figure}
\plotone{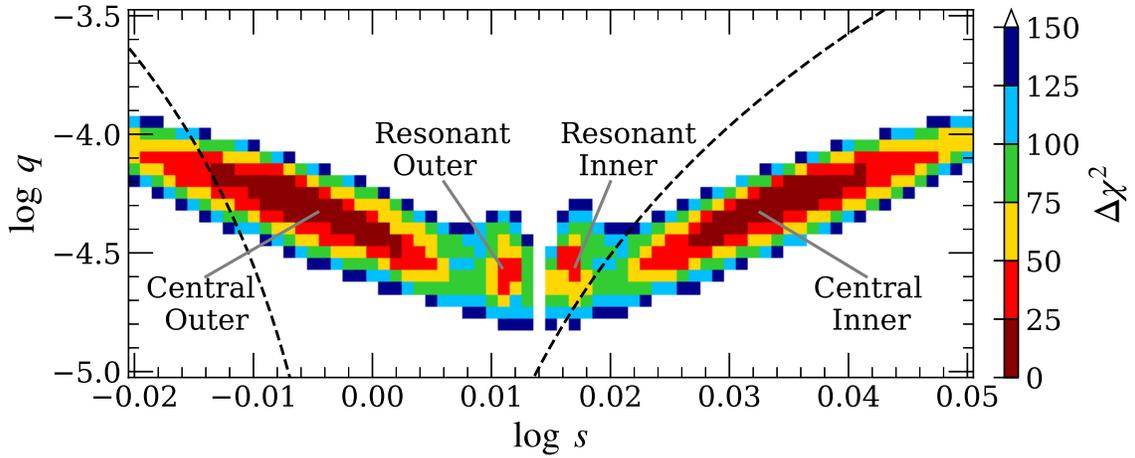}
\caption{Results of a ``grid search'' for solutions in which
$(s,q)$ are held fixed on a grid of values, while the remaining 5 parameters
  are allowed to vary.  The two ``central-caustic'' solutions that were
  previously discovered \citep{ob161195a,ob161195b} were recovered, but
  two additional ``resonant-caustic'' solutions were also discovered.  Full
  analysis of the combined data set (Table~\ref{tab:wz}) rules out these
  solutions.  The boundaries between the central-caustic and resonant-caustic
  regimes are indicated by dashed lines.
}
\label{fig:grid}
\end{figure}

\begin{figure}
\plotone{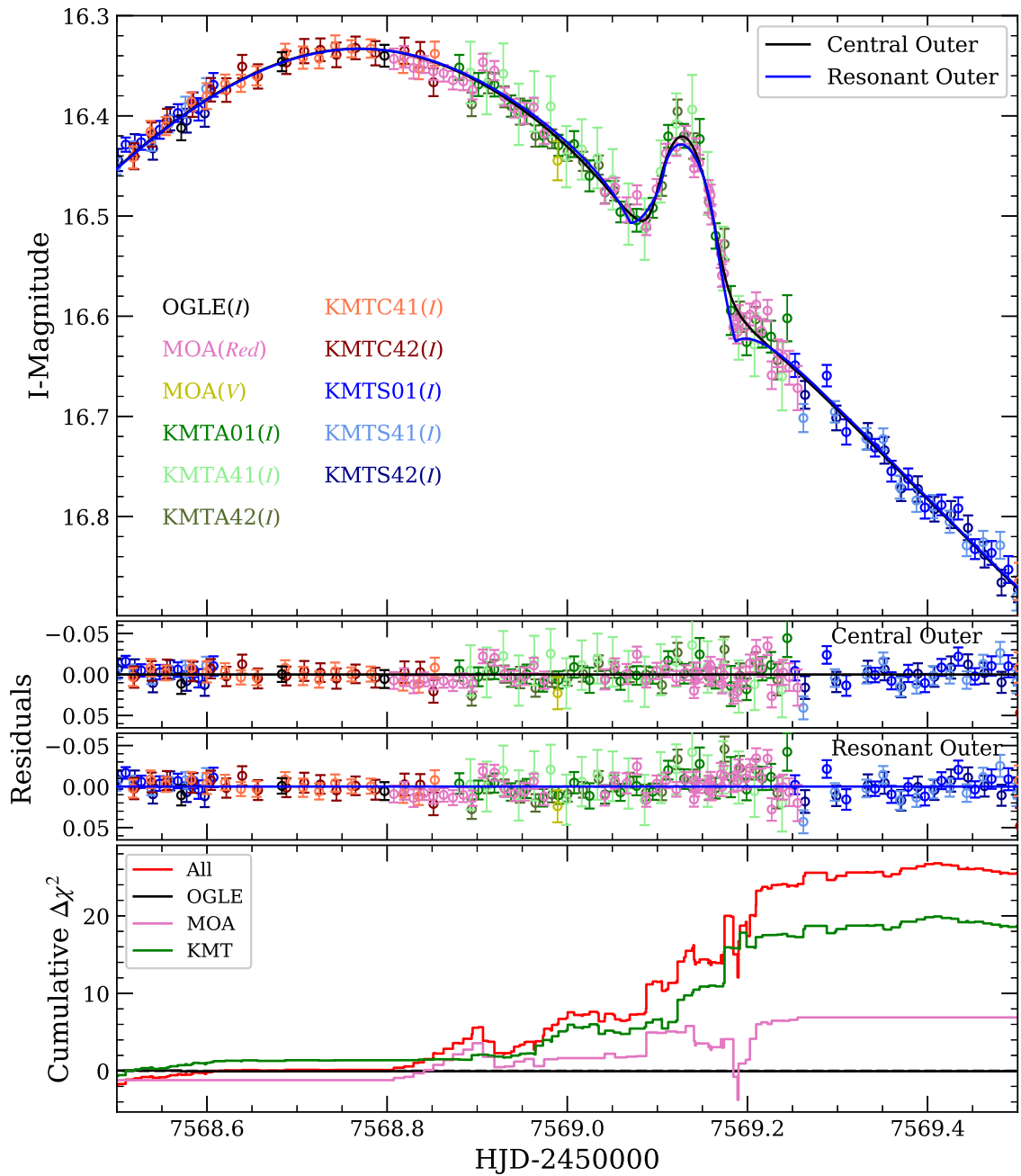}
\caption{Data (colored points) and two representative models of
  OGLE-2016-BLG-1195, a central-caustic model (black) that was originally
  found in the discovery papers \citep{ob161195a,ob161195b}, and a new
  resonant-caustic model.  The cumulative $\Delta\chi^2$ function shows that
  the new resonant model is strongly disfavored, primarily due to its failure
  to match the anomalous region, 7569.08--7569.22.  However, rejection
  of this model would have been less decisive based on the partial data sets
  in each discovery paper.  See also Figure~\ref{fig:caustic}.
}
\label{fig:lc}
\end{figure}

\begin{figure}
\plotone{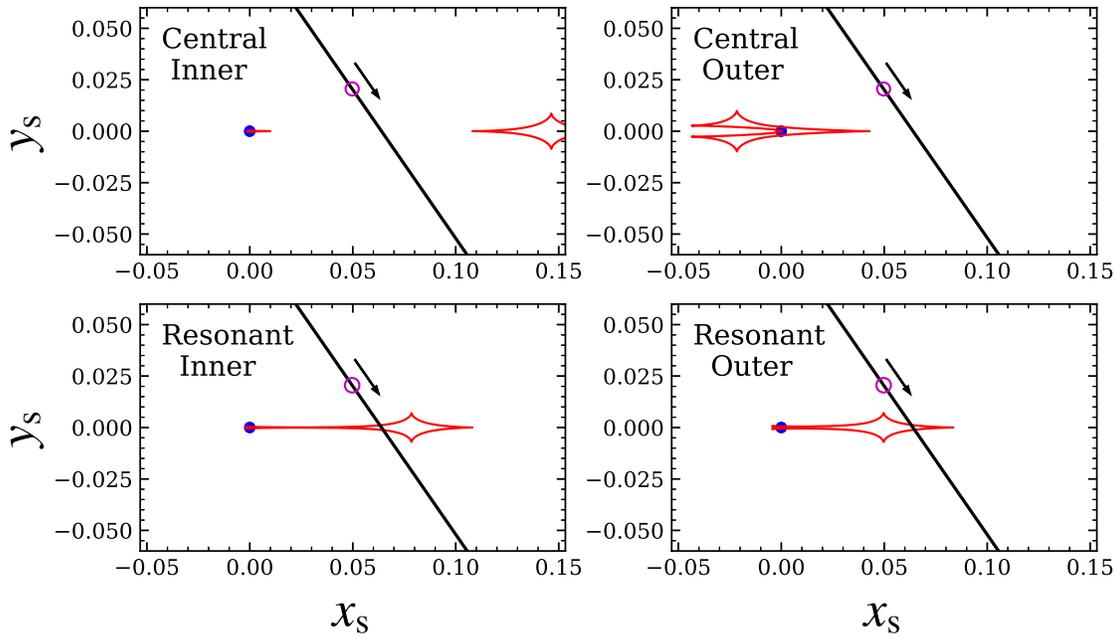}
\caption{Caustic geometries for the four solutions shown in
  Table~\ref{tab:wz} and illustrated in Figure~\ref{fig:lc}.  In the
  ``central caustic'' models, the ``bump'' in the light curve centered
  at 7569.13 is caused by the source passing over a ridge that extends from
  the central (or resonant) caustic, whereas in the ``resonant caustic''
  models, it is caused by the source passing over two closely-spaced caustics
  of the planetary wing of a resonant caustic.
}
\label{fig:caustic}
\end{figure}

There is one major and one minor feature of note about the new analysis.
The major feature is that
there are four solutions, including the two solutions found in the
original papers \citep{ob161195a,ob161195b}, plus two additional solutions
that were identified in our new grid search (Figure~\ref{fig:grid}).
In the original solutions, the bump peaking at 7569.13 in Figure~\ref{fig:lc}
is a major-image anomaly generated by
the source crossing a ridge that extends from the central (or resonant)
caustic in the direction of the planet.  In the additional solutions,
the source crosses two closely spaced caustics in the planetary wing of
a resonant caustic.  See Figure~\ref{fig:caustic}. This degeneracy was
discovered by \citet{kb211391} and was dubbed the ``central-resonant''
degeneracy by \citet{kb210171}.

Of particular relevance in the current context is that such a difference
in morphologies can significantly impact $t_*$.  This occurred for both of the
cases in \citet{kb211391}, (i.e.,
KMT-2021-BLG-1391 and KMT-2021-BLG-0253), as well as for one of the two cases
in \citet{kb210171} (i.e., KMT-2021-BLG-1689 but not KMT-2021-BLG-0171).
As \citet{kb211391} recount, this degeneracy
was discovered by a systematic effort to identify symmetries among multiple
solutions.  They hypothesized that because the degeneracy was found in 2 out
of 4 randomly chosen events, it could be common and therefore
should be checked for in archival events
with bump-type anomalies at relatively high magnification.  OGLE-2016-BLG-1195
is indeed a typical example of such an anomaly.

Table~\ref{tab:wz} shows that the additional solutions (labeled ``resonant'')
indeed have significantly different values of $t_*$ from the original solutions.
Specifically, the values are
$t_*=0.0317\pm 0.0017\,$day (central) versus
$t_*=0.0350\pm 0.0009\,$day (resonant).  This difference is far smaller
than the factor $\sim 2$ difference found by \citet{kb211391} for their
two cases.  Moreover, in the present context, the new solutions have
larger $t_*$ (so smaller $\mu_\rel$), which would only strengthen the tension,
not relieve it.  Finally, the resonant solutions are disfavored in the
combined analysis by $\Delta\chi^2=27$, which would normally lead to their
being reported but discounted in the final assessment of the planet's
characteristics.  Nevertheless, because the $\chi^2$ difference is 
divided between the two independent data sets
(see Figure~\ref{fig:lc}), it would have
been substantially smaller in the original analyses had these additional
solutions been recognized, and it could not have been so decisively
rejected\footnote{Note that there is a plotting error in Figure~2 of
\citet{ob161195a}, which makes it appear as though the MOA data would favor
the resonant-caustic models, whereas actually they disfavor these models.
That is, there appears to be one prominent point that is below the displayed
central-caustic model at 7569.08 and several more near 7569.22, i.e., exactly
the places that the resonant-caustic model predicts dips relative to the
central-caustic model.  The most likely explanation is that, in making these
plots, \citet{ob161195a} first calculated the flux residuals at time $t_i$
and observatory $j$ as $\Delta f_{i,j} = f_j(t_i) - (f_{S,j} A(t_i) + f_{B,j})$
and then plotted their data points at $A_j(t_i) = A(t_i) - \Delta f_{i,j}/f_{S,j}$.
However, these should have been plotted at
$A_j(t_i) = A(t_i) + \Delta f_{i,j}/f_{S,j}$.  See the residuals panel in
Figure~\ref{fig:lc} of the present paper for the correct plotting.}.
In particular, if the resonant solution had been recognized,
it could not have been ruled out based solely on the $\Delta\chi^2\sim 7$
difference from an analysis restricted to OGLE and MOA data.
Nevertheless, this modest $\Delta\chi^2$ combined with the phase-space
arguments of \citet{kb210171}, would (in a more modern context)
probably be enough to reject this solution.

The minor feature is that the best-fit value of $t_*$ is slightly
smaller than the one in Table~\ref{tab:combo} and also has a smaller
error bar: $t_*=0.0317\pm 0.0017\,$day versus the previous value of
$t_*=0.0324\pm 0.0019\,$day.  While this change is small compared to the
errors, it still must be taken into account to make the best estimate
of the discrepancy between the light-curve-based and AO-based proper-motion
measurements.

The net result of this investigation is that the light-curve-based proper
motion increased from $\mu_\rel=9.27 \pm 0.69\,\masyr$
(Equation~(\ref{eqn:mureleval})) to $\mu_\rel=9.52 \pm 0.69\,\masyr$.
Comparing to Equation~(\ref{eqn:helpm}) (and ignoring the correction from
heliocentric to geocentric), this implies a conflict of $3.9\,\sigma$
rather than $4.0\,\sigma$.  Thus our light-curve reanalysis confirms
the strong tension between the light-curve- and AO-based proper-motion
measurements.

\section{{Discussion}
  \label{sec:discussion}}

High-resolution imaging of the hosts of microlensing planets, which
was pioneered by \citet{ob05169bat} and \citet{ob05169ben} for Keck-AO
and {\it Hubble Space Telescope (HST)} imaging, respectively, has immense
scientific prospects.  As demonstrated by \citet{gould22}, this
approach can yield planetary and host masses for a statistically
complete sample of well over 100 planets, beginning at first AO light
on ELTs, probably about 2030.  There is no other technique that can do
this, in particular covering such a wide range of host and planet
masses, projected separations, and Galactic environments.

However, as also shown by \citet{gould22}, only very fragmentary
results are possible prior to ELT observations because only a small
subset of the full sample will be accessible to present-day
instruments.  Hence, the main scientific return of high-resolution
observations today is technical in nature, in effect, establishing and
refining the viability of the technique.  A good example is provided
by \citet{ob161195c}, who were able to detect and measure the 16:1
flux ratio of a neighbor at 54 mas, i.e., inside the Keck-AO
FWHM.  This, together with earlier related achievements by the same
group, greatly increases confidence in the method, a confidence that
will be sorely needed to gain the necessary observing time on ELTs in
a highly competitive environment.

Nevertheless, as also systematically analyzed by \citet{gould22}, the
technical issues arising from this method are not restricted to imaging
technology: false positives and false negatives must be driven down to
an acceptable level.  Hence, the techniques for identifying these
must not only be cataloged in theory, they must also be tested in practice.

In this paper, we have shown that the host identification reported by
\citet{ob161195c} raises three different red flags.  The most striking
of these, and also the one that \citet{gould22} argued would be the most
common indication of a false positive,
is the strong conflict between the AO-based proper-motion
measurement (Equation~(\ref{eqn:helpm})) and the one derived from the 
published light-curve analyses (Equation~(\ref{eqn:mureleval})).
The second red flag (argued by \citealt{gould22} to be less common)
is the conflict with a previous microlens-parallax measurement.
In addition to these two known red flags, we also found
that the \citet{ob161195c} host identification leads to a solution
that is kinematically disfavored relative to the one derived from the
{\it Spitzer} microlens-parallax measurement.

None of these red flags proves that this host identification is
incorrect. Together, however, they do imply that the result must be more
deeply investigated, and if these investigations do not lead to clear
rejection, then the identification must be confirmed by directly
measuring the proper motion of the putative host relative to the
source by additional late-time observations.

Prompted by these three red flags, we have conducted such an
investigation based on existing data.  We noted that the source
appears to be ``too bright'' relative to the flux values implied by
the light-curve analysis, something that was also previously noted by
the \citet{ob161195c} authors but not reported in their published
paper.  We considered various explanations for this discrepancy,
including one in which \citet{ob161195c} may have identified the wrong
``star'' (or stellar asterism) on which to conduct their analysis.  We
then found preliminary evidence in favor of this hypothesis.

An important conclusion of \citet{ob161195c} was that by ``disproving''
the {\it Spitzer}-parallax solution, they helped to demonstrate the
general unreliability of {\it Spitzer} parallax measurements.  While we
do not agree that the {\it Spitzer} measurement has in fact been ``disproved'',
we do agree that testing of {\it Spitzer} parallax measurements by
high-resolution imaging is extremely important.  \citet{gould22} cataloged
12 planetary events with {\it Spitzer} parallaxes (his Table 3), of which
5 have giant sources and lens-source relative proper-motion
measurements (his Figure 4).  Because
giant sources can require much longer wait times, these potential targets
may have to wait until well after ELT AO first light.  In such cases, the
{\it Spitzer}-based mass measurements will be essential to including
these planets in statistical studies.  Hence, testing the reliability of
{\it Spitzer}-based parallaxes on events with fainter sources, e.g.,
OGLE-2016-BLG-1195, is crucial for establishing the conditions under which
these giant-source planetary events can be included.  We note that new
{\it Spitzer} planets are still being discovered (so not yet cataloged by
\citealt{gould22}), including OGLE-2019-BLG-0679 \citep{2019subprime} and
OGLE-2017-BLG-1275 \citep{2017prime}, with the former having a giant source.
Thus, one of the technical goals of current high-resolution studies should
definitely be to test {\it Spitzer} parallaxes when feasible, in particular
focusing on those that, like OGLE-2016-BLG-1195, do not have giant-star sources.

\begin{acknowledgements}
This research has made use of the Keck Observatory Archive (KOA), which is operated by the W. M. Keck Observatory and the NASA Exoplanet Science Institute (NExScI), under contract with the National Aeronautics and Space Administration.
J.C.Y. acknowledges support from US NSF Grant No. AST-2108414.
Y.S. acknowledges support from BSF Grant No. 2020740.
J.Z. and W.Z. acknowledge support by the National Science Foundation of China (Grant No. 12133005).
W.Z. acknowledges the support from the Harvard-Smithsonian Center for Astrophysics through the CfA Fellowship.
\end{acknowledgements}

\appendix
\section{Position of the Source Star}
\label{sec:sourcepos}

We measure the position of the source relative to the frame of
field stars in its neighborhood.  Such a measurement can potentially
have two different applications.
First, it allows us to identify the star (or asterism)
on the Keck-AO images that contains the source and therefore should
be the target of detailed analysis.  Second, it could also permit a
measurement of the source proper motion relative to this frame.  Then,
in turn, the frame proper motion could be put on an absolute scale using
the {\it Gaia} sources within it.  Combining such a measurement of
$\bmu_{S,\hel}$ with a measurement of $\bmu_{\rel,\hel}$ from the imaging
would yield $\bmu_{L,\hel} = \bmu_{\rel,\hel} + \bmu_{S,\hel}$, which could
help clarify the nature of the host.

These two potential applications have very different accuracy requirements.
The surface density of stars (or asterisms) that are bright enough to
contain the $K\sim 17.25$ source is relatively low.  Hence, barring a
pathological pile-up of such stars, an accuracy of a few tens of mas should
be adequate.  In fact, there are only two ``stars'' in
the broad neighborhood of the source that are sufficiently bright to contain
the source, and these are separated by $\sim 175\,\mas$.  See Figure~1 of
\citet{ob161195c}.  Hence,
few-tens-of-mas accuracy is indeed enough.  On the other hand, in order to
significantly constrain $\bmu_L$, the accuracy of $\bmu_S$ should be
$\la\sigma_\mu\sim 3\,\masyr$, i.e., the known dispersion of bulge stars.
Hence, the position measurement should have an accuracy
$\la \sigma_\mu \Delta t\sim 12\,\masyr$, which is more demanding.

The steps toward making this measurement are basically standard but still
require some description.  The first step is to measure the source
position on the reference frame of the field stars in the pixel-coordinate
frame of the camera.  For this purpose, we start by making pyDIA \citep{pydia}
reductions of the
KMTC01 images.  The reference image is formed by stacking many good images.
It is aligned to and then subtracted from a series of images in which
the source is magnified.  The resulting difference images then basically
consist of an isolated point-spread function (PSF), whose position is
easy to measure\footnote{We note that under certain conditions, this method
can fail, even catastrophically.  However, in Appendix~\ref{sec:failure}, we
derive a general formula that describes such potential failures and show
that they do not apply to the present case.}.

\begin{figure}
\plotone{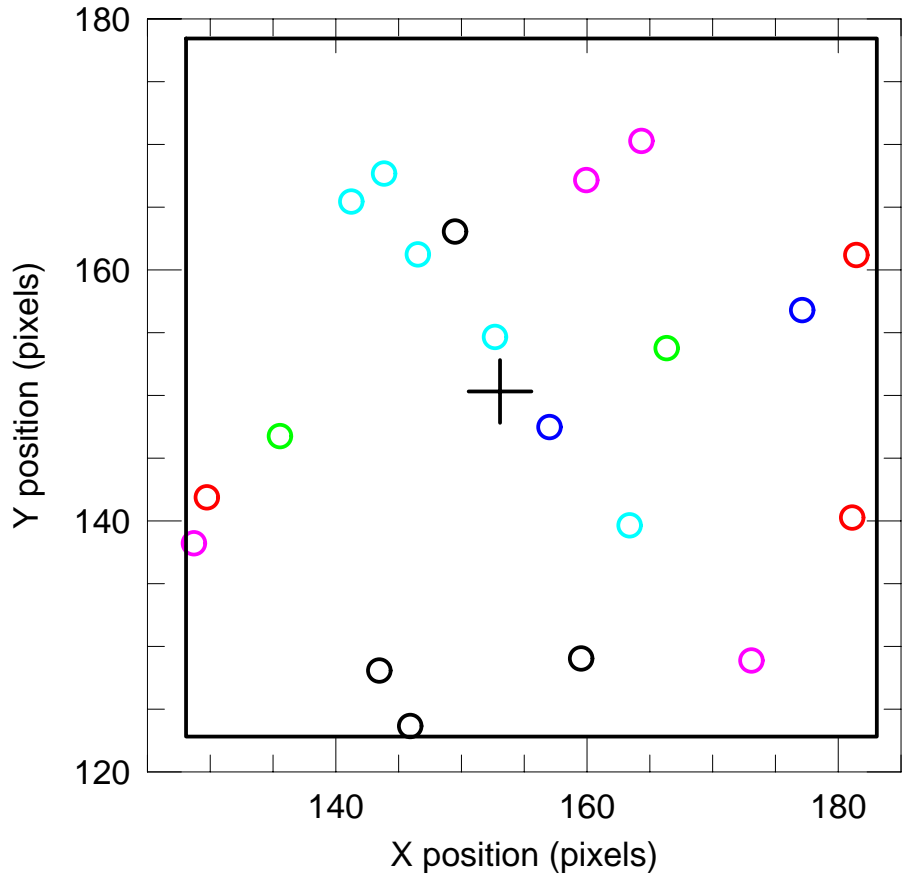}
\caption{Positions of the 20 brightest stars (in $I$ band) from
the KMTC01 pyDIA reduction that overlap the Keck OSIRIS
image (solid rectangle).  The star positions are in the original
reference-frame pixel coordinates.  The pixels are
approximately $0.4^{\prime\prime}$.  For ease of comparison to the Keck images,
the stars are color-coded according to $K_{\rm pseudo}\equiv I - (V-I)$ by
$K_{\rm pseudo}<(13,13.5,14,14.5,15,15.5)\rightarrow$
(black, red, green, cyan, blue, magenta).  The source position
$(X,Y)_{\rm source}= (153.074,150.317)$ is shown by a cross.
}
\label{fig:square}
\end{figure}

Next we repeat this procedure on three additional data sets: KMTC41, KMTS01
and KMTA01.  We align the images by cross-matching relatively bright, $I<17.5$
stars, making a 2-dimensional linear (6 parameter) transformation,
and iteratively rejecting
outliers.  Then, for each cross-matched star (and also for the microlensed
source position), we have 4 separate measurements, all on KMTC01 reference
system.  From these we find the mean position, scatter (standard deviation),
and standard error of the mean (s.e.m.).
In particular, for the microlensed source,
we find transformed positions
(153.0735,150.3169),
(153.0633,150.2900), 
(153.1118,150.3588), and
(153.1165,150.3539) for
KMTC01, KMTC41, KMTS01, and KMTA01, respectively.
This yields a mean and s.e.m. of
\begin{equation}
  (X,Y)_{S,\rm KMTC01} = (153.0913,150.3299) \pm (0.0134,0.0168)\ {\rm pixels},
  \label{eqn:source-position}
\end{equation}
corresponding to errors of $(5.4,6.7)\,$mas\footnote{Note that we include
KMTC01 in the pyDIA astrometric analysis even though it was excluded from
the photometric analysis in Section~\ref{sec:tens2}.  The astrometric analysis
depends on a relative handful of images, and we check by eye that the
contamination from the spike is low.  Finally we check that in the above
list of four measurements, KMTC01 contributes $\chi^2_{\rm KMTC01}=0.8$, compared
to an average value $\chi^2_{\rm ave}\equiv 1.5$, which is enforced by setting
$\chi^2_{\rm total}\equiv {\rm dof} = 6$ for 8 measurements.}.
In Table~\ref{tab:starlist},  we list the pixel positions and
instrumental magnitudes and colors of the 20 brightest stars lying in the
region of the Keck OSIRIS image.  We also list the pseudo-$K$-band magnitude,
$K_{\rm pseudo}\equiv I - (V-I)$ as a rough indicator of the relative
brightness expected in the $K$ band\footnote{This is because, first,
$(I-K)_0 \sim (V-I)_0$ over a broad range of stellar types \citep{bb88}.
If it were also the case that $E(I-K)\sim E(V-I)$, then $K_{\rm pseudo}$
would be a very good proxy for $K$.  In fact, for this line of sight, it
is roughly true that
$E(V-I)\sim E(I-K)$.  Even for other lines of sight, the great majority of
the stars in these images are in the bulge, and so suffer very similar
extinction.  To the extent that this is the case, the {\it relative}
brightness in $K$ is accurately predicted by $K_{\rm pseudo}$, making it
a valuable, even if imperfect, indicator of $K$.}.  These indicators are used in
Figure~\ref{fig:square} to color the points that represent these 20 stars,
which should enable easy comparison to the Keck $K$-band images.

Using these, together with the microlensed source position
in Equation~(\ref{eqn:source-position}),
it will be possible for any reader to
make his/her own estimate of the source position on the Keck image.

Next we proceed to make our own such measurements for the NIRC2 (2018)
and OSIRIS (2020) epochs.

First, we retrieved the Keck images via the Keck Observatory Archive
(KOA\footnote{https://www2.keck.hawaii.edu/koa/public/koa.php}).
For the NIRC2 2018 epoch we use 13 good images. 
For the OSIRIS 2020 epoch we use 23 good images.
For each image (in each epoch) we construct an astrometric
catalog using PSF fitting for the centroids of all sources detected in
the vicinity of the event, using the software package described in
\citet{ofek14} and Ofek et al. (in prep)
and specifically the astrometric tools
used in \citet{ofek19}.  The (23 or 13) catalogs are then aligned using
third-order polynomials.  The mean position and standard deviation is then
measured for each entry.  We then set the zero point of this
catalog as the measured position of the ``northwest star'', which we will
find below is the closest to the microlensed source.  Note that the
standard errors of the mean of this star are $(0.10,0.06)\,\mas$
and $(0.44,2.23)\,\mas$, for OSIRIS and NIRC2, respectively,
which are small compared to other errors in
the problem.  We express these offsets in arcsec.  Note that in the
KMT system, the first coordinate (in pixels) increases to the west,
while in the Keck system, the first coordinate (in arcseconds) increases
to the east.

Next, we identify a restricted subset of the KMT stars that have
{\it Gaia} counterparts with proper-motion measurements.  We exclude
from consideration any star with KMT position errors (s.e.m.) greater
than 10 mas.  In fact, all but one of these excluded stars have {\it Gaia}
entries but without proper-motion measurements, likely due to crowding
and/or unresolved sources.  For each such star, we find (if present) the OSIRIS
and NIRC2 counterparts.  Table~\ref{tab:gaialist} lists the resulting 13 stars.
Column 1 is a cross reference to Table~\ref{tab:starlist}.
Columns 2 and 3 give the {\it Gaia}
proper motions as well as their
errors (in the second row). Column 4 gives the {\it Gaia} RUWE indicator.
Columns (5,6) and (7,8) give the OSIRIS and NIRC2 positions, with
their standard deviations shown in the second row.  Note that for four of the
$2\times 13=26$
cases, there is no measurement (hence, no entry).  Thus, there are potentially
10 cross matches for OSIRIS and 12 for NIRC2.  However, we find that the
brightest three NIRC2 stars are seriously saturated, with bagel-morphology
images.  Corresponding to this, their standard deviations are dramatically
larger than those of the fainter NIRC2 stars.  Hence, we exclude these,
leaving 9 NIRC2 stars.


Next we propagate the KMT positions forward using the {\it Gaia} proper
motions by 2.10 and 4.12 years for NIRC2 and OSIRIS, respectively.
For this purpose we subtract an estimate of the mean proper motion
of the bulge, $\bmu_{\rm bulge}(E,N) = (-3.00,-5.20)\,\masyr$, from each
{\it Gaia} proper motion.  That is, ultimately we will be measuring
the offset between the positions of Keck stars in the 2018/2020 bulge frame
from the position of the microlensed source in the 2016 bulge frame.

We multiply the reported {\it Gaia} errors by 1.5 to take account of the
general difficulty of proper-motion measurements in the bulge and then
add these in quadrature to the KMT position errors.  We also include
the s.e.m.\ of the Keck measurements, but these are generally too small
to matter.

We then carry out a second order (quadratic) transformation from the KMT frame
to the Keck frames and thus derive the positional offsets of the
``northwest star'' from the nominal position of the microlensed source.  We
find offsets (``northwest star'' minus microlensed source) of
$\Delta(E,N)_{2018} = (-26.8,-1.6)\pm (1.2,2.2)\,\mas$
with $\chi^2/{\rm dof}=90/6$ and
$\Delta(E,N)_{2020} = (-46.3,-4.5)\pm (1.9,3.2)\,\mas$
with $\chi^2/{\rm dof}=108/8$, for NIRC2 and OSIRIS, respectively.

Clearly, our formalism has not captured all sources of error.  One
potential cause of additional errors is that the KMT and Keck images have
substantially different resolutions and bandpasses, so that each can be
affected in different ways by stars that are uncataloged, either because
they are too faint or are buried within the PSFs of cataloged stars.
There could be others.  We adopt the simple expedient of renormalizing the
final error bars by $\sqrt{\chi^2/{\rm dof}}$ to obtain
\begin{equation}
\Delta(E,N)_{2018} = (-26.8,-1.6)\pm (4.6,8.5)\,\mas,
\qquad
\Delta(E,N)_{2020} = (-46.3,-4.5)\pm (7.0,11.8)\,\mas
\label{eqn:nw-offset}
\end{equation}

\begin{figure}
\plotone{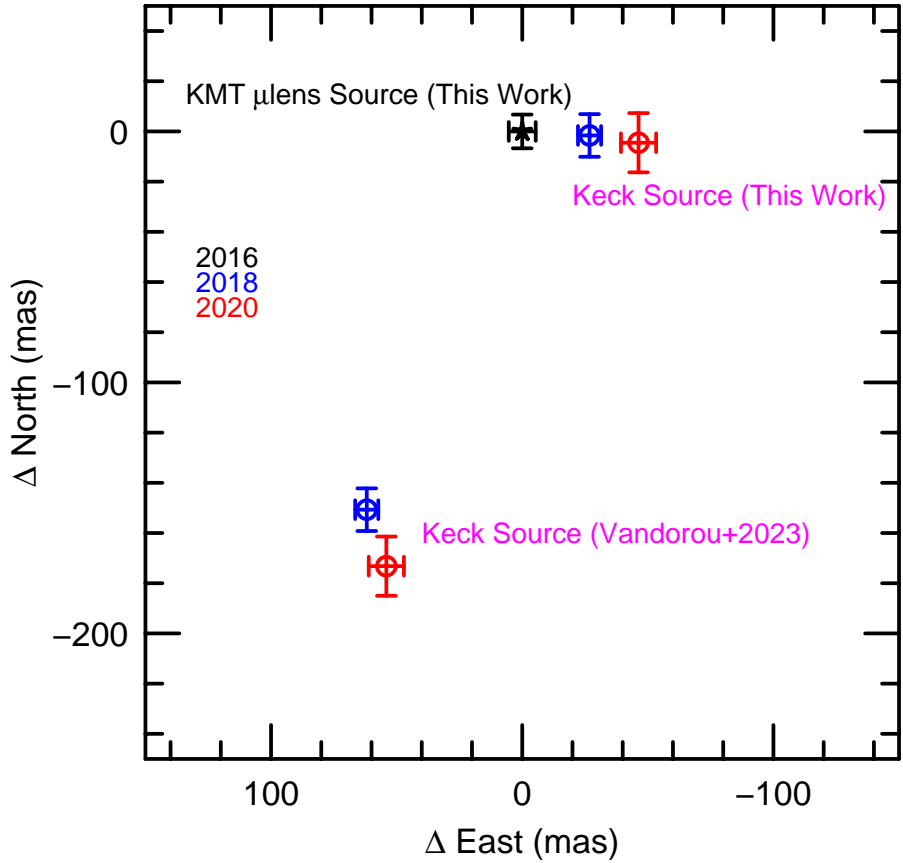}
\caption{Offsets in positions of two nearby stars relative to that of
  microlensed source (black) as determined from a difference imaging analysis
  of KMT data from the microlensing event in 2016.
  The two stars lie in the northwest and southeast quadrants
  of the zoom of Figure~1 from \citet{ob161195c}.  The analysis of the
  present work yields the offsets of these two stars at two epochs, i.e.,
  2018 (blue, NIRC2) and 2020 (red, OSIRIS).  The northwest star is much
  closer to the microlensed source, suggesting that the two are associated.
  By contrast, \citet{ob161195c} assumed that the southeast star was
  associated with the microlensing event.  Note that the black error bars
  reflect the precision of centroiding the source within the KMT difference
  images, whereas the blue and red error bars reflect the precision of
  transforming from the Keck images to the KMT reference image.
}
\label{fig:offset}
\end{figure}

To get a further handle on the errors, we repeat these calculation by
eliminating either the largest or two-largest outliers.
In 7 of the 8 cases, (2 epochs)$\times$(2 eliminations)$\times$(2 components),
we find that the changes are $< 10\,\mas$, while in one case
(NIRC2, 2 eliminations, declination), the change is 24 mas.
Hence, we estimate that the true errors are of order 15 mas.
Nevertheless, for purposes of
display in Figure~\ref{fig:offset}, we show the formal errors of
Equation~(\ref{eqn:nw-offset}).  This Figure also shows the microlensed
source position and the position of the ``southeast star'', i.e., the
one identified by \citet{ob161195c} as the microlensed source.  Recall from
the discussion above that the Keck star positions are in the bulge frame at
their respective epochs, while the microlensed source position is in the
bulge frame from 2016.  Hence, one possible reason that the microlensed
source is displaced from the northwest star is that the latter moved toward
the west in the intervening years.  However, the two Keck positions
are actually consistent within the errors.  Hence it is also possible
that the northwest star has moved very little in the bulge frame, while the
microlensed source has remained within the Keck PSF of this bright star.

\begin{figure}
\plotone{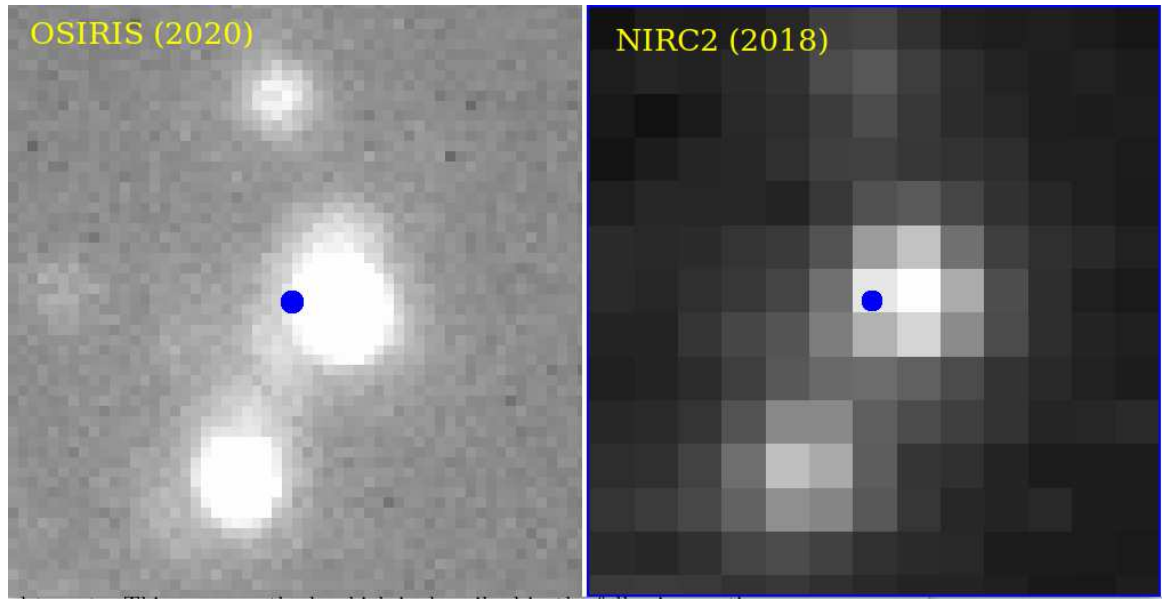}
\caption{Keck images of the region surrounding OGLE-2016-BLG-1195 from
  OSIRIS 2020 (left) and NIRC2 2018 (right).  In each case the position of the
  microlensed source (in the 2016 bulge frame) is superposed as a blue
  circle.  The pixel sizes are, respectively, 10 and 40 mas.  North is
  up and East is left. \citet{ob161195c} identified the southeast
  star (or asterism) as the location of the microlensing event, whereas
  our analysis suggests that this is more likely to be the northwest star.
}
\label{fig:keckimage}
\end{figure}

Figure~\ref{fig:keckimage} shows the Keck images from the two epochs
(NIRC2 2018 and OSIRIS 2020) with the (2016 bulge-frame) position of the
microlensed source superposed as a blue circle.

Currently, the balance of evidence favors that the microlensed source
is associated with the northwest star, rather than the southeast star
that was identified by \citet{ob161195c}.  However, in our view, it would
be premature to claim this as a fact.  First, the evidence that we have
presented must be weighed against the evidence (based on completely independent
microlensing survey images) that led \citet{ob161195c} to the conclusion
that their southeast-star identification was correct.  Second, as we have
discussed, the non-linear transformation between the KMT and Keck frames
are limited by the number of stars.  In fact, we lacked enough stars
to even make the standard third-order (cubic) transformation.  One way
to resolve this would be to take an {\it HST} image in the $I$ band.
This would require only a linear transformation.  Moreover, it would
allow direct comparison with the microlensing photometry (also in the $I$
band), and so permit a more precise estimate of the excess flux.

However, perhaps the simplest approach would be to wait several more
years and then take a Keck or {\it HST} image.  For example, in 2025,
the source and lens will be separated by $\sim 90\,\mas$.  If the
{\it Spitzer} parallax is basically correct, then the host (near the
star/BD boundary) will
probably still not be visible.  But if the host has a more typical
mass, as argued by \citet{ob161195c}, then it will be separately resolved
in these images, probably near the northwest star, but likely visible
in either case.

For the present, the main result of this Appendix, is to reinforce
the need for a cautious approach to the identification of the host
of OGLE-2016-BLG-1195.

\section{Failure Mode of Difference-Image Astrometry}
\label{sec:failure}

The goal of difference-image astrometry is to locate the source
position on the seeing-limited reference image, with the ultimate aim of
comparing this reference-plus-source image to late-time high-resolution
images.  In the simplified presentation above, we implicitly assumed
that nothing had moved between the epochs of the images entering the
reference image and those of the magnified images.  However, for a variety
of reasons, including just convenience, the two may in principle
be separated by some interval $\Delta t$, which could be either
positive or negative.  For completeness, we note that this could, in
principle, lead to difficulties in aligning the reference and magnified
images due to random motions relative to the bulk motion of the frame,
but in practice this is almost never the case.  Hence, we
ignore this issue here.  We further note that in most cases, the
astrometric error of the source position is dominated by scintillation
noise, rather than photon noise.  However, this scintillation noise is
automatically accounted for in our approach (which is often adopted) of
estimating the astrometric errors from the scatter of multiple measurements.

After the magnified images (for simplicity, just called ``image'' ${\bf I}$)
are photometrically and astrometrically aligned to the reference image
${\bf R}$, and then ${\bf R}$ is convolved to match the PSF of ${\bf I}$,
a difference image ${\bf D} = {\bf I} - {\bf R}$ is formed by subtracting the
second from the first.  In the approximation that nothing has moved
between the construction of ${\bf R}$ and ${\bf I}$,
${\bf D} = f_S(A-1){\bf P}_S$,
where $f_S$ is the source flux, $A$ is the magnification,
and ${\bf P}_S$ is a normalized PSF whose
centroid is at the location of the source $\btheta_S$.  Then, in this
approximation,
\begin{equation}
\btheta_S = {\bf c}({\bf P}_S) = {{\bf c}({\bf D}_S)\over (A-1)f_S}
\label{eqn:centroid}  
\end{equation}
where ${\bf c}$ is the centroid operator that, in effect, just sums over the
pixel values in ${\bf P}_S$ (of ${\bf D}$)\footnote{In practice, the centroid
is often found by maximizing the likelihood of a fit to ${\bf D}_S$ of models
composed of a PSF at various positions plus a background term.  Provided
that the pixel counts are dominated by sources well within the PSF, this
procedure has the same expectation but smaller errors.  However, the
process cannot be represented mathematically as simply as the
${\bf c}$ operator in Equation~(\ref{eqn:centroid}).}.

However, if either the microlensed source or the blended light, $f_B$,
lying within the PSF (or both) have moved, then the subtraction operation
will yield ${\bf P}_{\rm diff} = ({\bf I} - {\bf R})/[f_S(A-1)]$, whose
centroid is displaced from $\btheta_S$ by
\begin{equation}
  \Delta\btheta = {\bf c}({\bf P}_{\rm diff}) - \btheta_S =
{{\bf c}({\bf I}-{\bf R})\over (A-1)f_S} - \btheta_S
\label{eqn:displace}  
\end{equation}
In general, the blended light can take complex forms.  However, as long
as both the source and the components of the blended light remain well
within the FWHM of the PSF, one can ignore these complexities.  This
is the case we consider here.  It is not difficult, in principle, to consider
blends that, e.g., straddle the PSF.  However, such a more general treatment
would take us too far afield.  Then, Equation~(\ref{eqn:displace}) can be
evaluated,
\begin{equation}
  \Delta\btheta = 
 {(f_S A\btheta_{S,1} + f_B\btheta_{B,1}) - (f_S\btheta_{S,0} + f_B\btheta_{B,0})
 \over(A-1)f_S} - \btheta_{S,0},
\label{eqn:displace2}  
\end{equation}
where the subscripts ``0'' and ``1'' refer to the reference epoch and the
magnified epoch, respectively.  Taking note that
$\btheta_{S,1} = \btheta_{S,0} + \bmu_S\Delta t$ and
$\btheta_{B,1} = \btheta_{B,0} + \bmu_B\Delta t$, Equation~(\ref{eqn:displace2})
simplifies to
\begin{equation}
  \Delta\btheta = \biggl[{A\over A-1}\bmu_S + {f_B\over (A-1)f_S}\bmu_B\biggr]
  \Delta t .
\label{eqn:displace3}  
\end{equation}

The first point to note about Equation~(\ref{eqn:displace3}) is that if
$\Delta t$ is small, then this effect is negligible.  This is the case
for our measurement because the reference image is constructed from
epochs of the same year as the event and on either side of the peak.
Specifically, the mean epoch of the reference image is displaced from
$t_0$ by $\Delta t=-0.03\,$yr.
However, high-quality reference images are often used 
to carry out photometry for many years.  This is feasible because
the photometric error induced (for an isolated source and in the
approximation of a Gaussian PSF) is just
$\delta I = \log(32)(\delta \theta/{\rm FWHM})^2$, which is  tiny provided
that the source has moved $\delta\theta = \mu_S\Delta t \ll {\rm FWHM}$.

However, the astrometric effects can be much larger.  For example,
for very low-magnification events, which includes some of the important
class of giant-source free-floating planet (FFP) candidates (e.g.,
OGLE-2012-BLG-1323, \citealt{ob121323}), it is possible for $A/(A-1)\ga 10$,
so that even with typical $\mu_S\sim 4\,\masyr$ and a modest time offset
from the reference image, $\Delta t\sim 5\,$yr, the error in the source
position could be $\Delta \theta_S\ga 200\,\mas$, which could dramatically
impact the science interpretation.  See, e.g., \citet{gould14}.  Similarly,
such typical motions of a giant-star blend that was a factor 10 brighter than
the magnified source could create similar artificial offsets.  For example,
such a blend could lie at 200 mas, so well inside the seeing-limited PSF,
but easily resolved in AO images of next-generation ELTs.

In most cases, these problems can be avoided simply by constructing
reference images from the same year as the peak magnification of the
event.  However, this may be difficult in some cases, while in others,
the specific application might require even greater care to assure that
$\Delta t$ is as close to zero as possible.

%


 \begin{deluxetable}{lrrrrrrr}                                  
 \tablecolumns{8} \tablewidth{0pc}                              
 \tablecaption{\textsc{Comparison and Combination of Two Fits}} 
 \tablehead{\colhead{Parameter} &                               
\colhead{B+2017} &                                              
\colhead{$\sigma$(B+)} &                                        
\colhead{S+2017} &                                              
\colhead{$\sigma$(S+)} &                                        
\colhead{$\Delta/\sigma$} &                                     
\colhead{Comb.} &                                               
\colhead{$\sigma$(C)} }                                         
 \startdata                                                     
$t_0 - 2457560$    &  8.7716 &  0.0020 &  8.7693 &  0.0013 & $  0.9432$ &  8.7700 &  0.0011\\
$t_\eff$ (day)     &  0.5225 &  0.0069 &  0.5296 &  0.0046 & $ -0.8605$ &  0.5274 &  0.0038\\
$t_\e$ (day)       & 10.1850 &  0.2550 &  9.9600 &  0.1100 & $  0.8102$ &  9.9953 &  0.1010\\
$t_*$ (day)        &  0.0337 &  0.0022 &  0.0286 &  0.0037 & $  1.1730$ &  0.0324 &  0.0019\\
$t_q$ (hr)         &  0.0103 &  0.0016 &  0.0133 &  0.0018 & $ -1.2249$ &  0.0117 &  0.0012\\
$\alpha$ (deg)     & 55.2850 &  0.2600 & 55.4988 &  0.1300 & $ -0.7353$ & 55.4560 &  0.1163\\
$s_{\rm inner}$    &  1.0698 &  0.0078 &  1.0858 &  0.0078 & $ -1.4505$ &  1.0778 &  0.0055\\
$s_{\rm outer}$    &  0.9957 &  0.0073 &  0.9839 &  0.0072 & $  1.1508$ &  0.9897 &  0.0051\\
 \enddata                                                       
\tablecomments{B+=Bond+2017, S+=Shvartzvald+2017, C=Combined}   
 \label{tab:combo}                                              
 \end{deluxetable}

\begin{table*}
    \renewcommand\arraystretch{1.25}
    \centering
    \caption{Static 2L1S models for OGLE-2016-BLG-1195 (OGLE + MOA + KMT-new)}
    \begin{tabular}{c|c|c|c|c}
    \hline
    \hline
    Parameters &  \multicolumn{2}{c|}{Central} & \multicolumn{2}{c}{Resonant} \\ 
      & \multicolumn{1}{c}{Central Inner} & Central Outer & \multicolumn{1}{c}{Resonant Inner} & Resonant Outer \\
    \hline
    $\chi^2/\rm dof$  
    & $29461.4/30470$ & $29459.6/30470$ 
    & $29487.3/30470$ & $29486.6/30470$ \\
    \hline 
    $t_{0}-7560$ 
    & $8.7707 \pm 0.0010$ & $8.7706 \pm 0.0009$ 
    & $8.7707 \pm 0.0009$ & $8.7707 \pm 0.0009$ \\
    $u_0$  
    & $0.0528 \pm 0.0007$ & $0.0528 \pm 0.0007$ 
    & $0.0530 \pm 0.0007$ & $0.0531 \pm 0.0007$ \\  
    $t_\e$ (days) 
    & $9.91 \pm 0.11$ & $9.91 \pm 0.11$ 
    & $9.89 \pm 0.11$ & $9.88 \pm 0.10$ \\  
    $\rho (10^{-3})$ 
    & $3.20 \pm 0.18$ & $3.20 \pm 0.17$ 
    & $3.52 \pm 0.09$ & $3.56 \pm 0.10$ \\
    $\alpha$ (degree) 
    & $55.25 \pm 0.11$ & $55.23 \pm 0.11$ 
    & $55.28 \pm 0.11$ & $55.31 \pm 0.11$ \\  
    $s$  
    & $1.0763 \pm 0.0049$ & $0.9911 \pm 0.0044$ 
    & $1.0402 \pm 0.0006$ & $1.0254 \pm 0.0007$ \\
    $q (10^{-5})$ 
    & $4.62 \pm 0.44$ & $4.60 \pm 0.43$ 
    & $2.75 \pm 0.13$ & $2.77 \pm 0.15$ \\
    $\log q$ 
    & $-4.337 \pm 0.042$ & $-4.339 \pm 0.041$ 
    & $-4.561 \pm 0.021$ & $-4.559 \pm 0.023$ \\
    $t_\eff$ (days) 
    & $0.5237 \pm 0.0015$ & $0.5235 \pm 0.0015$
    & $0.5241 \pm 0.0016$ & $0.5246 \pm 0.0016$  \\   
    $t_*$ (days) 
    & $0.0317 \pm 0.0017$ & $0.0317 \pm 0.0016$  
    & $0.0348 \pm 0.0009$ & $0.0352 \pm 0.0009$  \\ 
    $t_q$ (hrs) 
    & $0.0110 \pm 0.0010$ & $0.0110 \pm 0.0010$  
    & $0.0065 \pm 0.0003$ & $0.0066 \pm 0.0003$  \\
    $f_{\rm S, OGLE}$ 
    & $0.2304 \pm 0.0030$ & $0.2304 \pm 0.0030$ 
    & $0.2311 \pm 0.0031$ & $0.2315 \pm 0.0029$ \\ 
    \hline
    \hline
    \end{tabular}
    \label{tab:wz}
\end{table*}

 \begin{deluxetable}{rrrrrrrrrr}
 \tablecolumns{10} \tablewidth{0pc}
 \tablecaption{\textsc{20 Brightest KMTC01 Stars}}
 \tablehead{\colhead{Star} &
 \colhead{$\Delta$E($^{\prime\prime}$)} &
 \colhead{$\Delta$N($^{\prime\prime}$)} &
 \colhead{$X$(pixel)} &
 \colhead{$Y$(pixel)} &
 \colhead{$\sigma(X)$} &
 \colhead{$\sigma(Y)$} &
 \colhead{$I$} &
 \colhead{$V-I$} &
 \colhead{$K_{\rm pseudo}$}}
 \startdata
 1 & $  3.85$ & $ -8.90$ & 143.4649 & 128.0831 &   0.0094 &   0.0064 &   13.85 &    2.68 &   11.16\\
 2 & $  2.87$ & $-10.66$ & 145.9149 & 123.6762 &   0.0074 &   0.0123 &   14.64 &    3.26 &   11.38\\
 3 & $  1.44$ & $  5.08$ & 149.4823 & 163.0342 &   0.0077 &   0.0124 &   15.55 &    2.82 &   12.73\\
 4 & $ -2.58$ & $ -8.52$ & 159.5384 & 129.0413 &   0.0039 &   0.0009 &   15.61 &    3.10 &   12.51\\
 5 & $-11.20$ & $ -4.03$ & 181.0989 & 140.2636 &   0.0586 &   0.0264 &   15.89 &    2.46 &   13.43\\
 6 & $  9.34$ & $ -3.38$ & 129.7343 & 141.8794 &   0.0020 &   0.0032 &   15.96 &    2.60 &   13.36\\
 7 & $-11.34$ & $  4.35$ & 181.4430 & 161.2108 &   0.0042 &   0.0074 &   15.98 &    2.60 &   13.38\\
 8 & $  2.63$ & $  4.38$ & 146.5206 & 161.2719 &   0.0073 &   0.0170 &   16.04 &    1.81 &   14.23\\
 9 & $  7.01$ & $ -1.43$ & 135.5583 & 146.7614 &   0.0044 &   0.0024 &   16.29 &    2.41 &   13.88\\
10 & $ -5.31$ & $  1.37$ & 166.3558 & 153.7593 &   0.0060 &   0.0065 &   16.34 &    2.45 &   13.89\\
11 & $  3.69$ & $  6.95$ & 143.8606 & 167.7049 &   0.0086 &   0.0165 &   16.59 &    2.27 &   14.32\\
12 & $  0.17$ & $  1.74$ & 152.6748 & 154.6741 &   0.0027 &   0.0155 &   16.66 &    2.35 &   14.30\\
13 & $  4.75$ & $  6.03$ & 141.2199 & 165.4033 &   0.0083 &   0.0267 &   16.72 &    2.52 &   14.21\\
14 & $ -4.12$ & $ -4.27$ & 163.3926 & 139.6519 &   0.0033 &   0.0066 &   17.04 &    2.85 &   14.19\\
15 & $ -1.57$ & $ -1.13$ & 157.0192 & 147.5050 &   0.0121 &   0.0160 &   17.21 &    2.45 &   14.77\\
16 & $ -9.61$ & $  2.57$ & 177.1237 & 156.7633 &   0.0388 &   0.0146 &   17.26 &    2.47 &   14.79\\
17 & $ -4.49$ & $  7.99$ & 164.3092 & 170.3038 &   0.0108 &   0.0095 &   17.31 &    1.85 &   15.46\\
18 & $  9.71$ & $ -4.78$ & 128.8218 & 138.3825 &   0.0485 &   0.0667 &   17.61 &    2.36 &   15.25\\
19 & $ -8.00$ & $ -8.57$ & 173.0839 & 128.8976 &   0.0022 &   0.0273 &   17.64 &    1.97 &   15.66\\
20 & $ -2.77$ & $  6.73$ & 160.0079 & 167.1636 &   0.0254 &   0.0040 &   17.75 &    2.64 &   15.11\\
 \enddata
 \tablecomments{$(X,Y)_{\rm source}=(153.0913,150.3299)\pm (0.0116,0.0141)$}
 \label{tab:starlist}
 \end{deluxetable}

 \begin{deluxetable}{rrrrrrrr}
 \tablecolumns{8} \tablewidth{0pc}
 \tablecaption{\textsc{13 KMT-{\it Gaia} Matches}}
 \tablehead{\colhead{Star} &
 \colhead{$\mu_\alpha$} &
 \colhead{$\mu_\delta$} &
 \colhead{RUWE} &
 \colhead{$\Delta$E(OS)} &
 \colhead{$\Delta$N(OS)} &
 \colhead{$\Delta$E(N2)} &
 \colhead{$\Delta$N(N2)}}
 \startdata
 1 &$   -1.36$ &$   -0.88$ &1.02 &           &           &           &          \\
   &      0.05 &      0.03 &     &           &           &           &          \\
 \hline
 2 &$   -4.93$ &$   -5.97$ &0.81 &           &           &$  2.5054$ &$-10.6576$\\
   &      0.11 &      0.07 &     &           &           &   0.0393  &   0.0336 \\
 \hline
 3 &$   -8.82$ &$   -3.92$ &1.19 &$  1.4265$ &$  4.9511$ &$  1.4354$ &$  4.9147$\\
   &      0.16 &      0.10 &     &   0.0005  &   0.0005  &   0.0328  &   0.0202 \\
 \hline
 4 &$   -3.45$ &$   -5.70$ &1.06 &$ -2.7065$ &$ -8.2932$ &$ -2.7759$ &$ -8.3684$\\
   &      0.12 &      0.07 &     &   0.0003  &   0.0008  &   0.0165  &   0.0222 \\
 \hline
 6 &$    2.33$ &$   -1.25$ &1.09 &$  9.2114$ &$ -3.4319$ &$  9.2014$ &$ -3.4270$\\
   &      0.15 &      0.11 &     &   0.0012  &   0.0005  &   0.0078  &   0.0043 \\
 \hline
 7 &$   -4.66$ &$  -10.95$ &1.01 &           &           &$-11.4346$ &$  4.4242$\\
   &      0.15 &      0.10 &     &           &           &   0.0030  &   0.0020 \\
 \hline
 9 &$   -2.16$ &$   -7.47$ &1.20 &$  6.9454$ &$ -1.5280$ &$  6.9400$ &$ -1.5243$\\
   &      0.19 &      0.13 &     &   0.0004  &   0.0003  &   0.0014  &   0.0024 \\
 \hline
10 &$   -5.20$ &$    0.82$ &1.19 &$ -5.3049$ &$  1.4491$ &$ -5.3170$ &$  1.4445$\\
   &      0.22 &      0.13 &     &   0.0004  &   0.0002  &   0.0016  &   0.0020 \\
 \hline
11 &$   -4.76$ &$   -3.20$ &1.56 &$  3.7042$ &$  6.7277$ &$  3.7029$ &$  6.7341$\\
   &      0.28 &      0.18 &     &   0.0004  &   0.0003  &   0.0009  &   0.0022 \\
 \hline
12 &$   -3.00$ &$   -5.05$ &1.18 &$  0.2190$ &$  1.7090$ &$  0.2019$ &$  1.7090$\\
   &      0.27 &      0.17 &     &   0.0004  &   0.0003  &   0.0034  &   0.0048 \\
 \hline
14 &$   -8.23$ &$   -9.56$ &1.28 &$ -4.1584$ &$ -4.1288$ &$ -4.1599$ &$ -4.1144$\\
   &      0.36 &      0.22 &     &   0.0005  &   0.0002  &   0.0011  &   0.0013 \\
 \hline
15 &$    0.42$ &$   -5.05$ &1.28 &$ -1.5333$ &$ -1.1035$ &$ -1.5521$ &$ -1.0972$\\
   &      0.36 &      0.23 &     &   0.0003  &   0.0002  &   0.0012  &   0.0015 \\
 \hline
17 &$    2.08$ &$   -6.53$ &0.94 &$ -4.4552$ &$  7.8875$ &$ -4.4669$ &$  7.8898$\\
   &      0.26 &      0.15 &     &   0.0004  &   0.0004  &   0.0012  &   0.0015 \\
 \enddata
 \tablecomments{units: proper motions ($\masyr$);  offsets ($^{\prime\prime}$)}
 \label{tab:gaialist}
 \end{deluxetable}

\clearpage

\end{document}